\pdfoutput=1

\documentclass[11pt]{article}

\usepackage{acl}

\usepackage[T1]{fontenc}
\usepackage[utf8]{inputenc} 
\usepackage{textcomp}
\usepackage{times}
\usepackage{latexsym}
\usepackage[T1]{fontenc}
\usepackage{microtype}
\usepackage{inconsolata}
\usepackage{graphicx} 
\usepackage{booktabs}
\usepackage{array}
\usepackage{tabularx}
\usepackage{amsmath}
\usepackage{amsfonts}
\usepackage{amssymb}
\usepackage{multirow}

\usepackage{subcaption} 
\usepackage{pgfplots}  
\pgfplotsset{compat=1.17}
\usepackage{tikz}
\usetikzlibrary{positioning, shapes.misc}

\title{Actors, Frames and Arguments: A Multi-Decade Computational Analysis of Climate Discourse in Financial News using Large Language Models}

\author{
  \textbf{Ruiran Su}$^{1}$ \quad \textbf{Janet B. Pierrehumbert}$^{1}$ \quad \textbf{Markus Leippold}$^{2}$ \\
  \\
  $^{1}$Department of Engineering Science, University of Oxford, Oxford, United Kingdom \\
  $^{2}$Department of Finance, University of Zurich, Zurich, Switzerland \\
  \\
  \texttt{\{ruiran.su, janet.pierrehumbert\}@eng.ox.ac.uk}, \texttt{markus.leippold@bf.uzh.ch}
}
\begin{document}
\maketitle

\begin{abstract}
Financial news media shapes trillion-dollar climate investment decisions, yet discourse in this elite domain remains underexplored. We analyze two decades of climate-related articles (2000–2023) from Dow Jones Newswire using an Actor–Frame–Argument (AFA) pipeline that extracts who speaks, how issues are framed, and which arguments are deployed. We validate extractions against 2,000 human-annotated articles using a Decompositional Verification Framework that evaluates completeness, faithfulness, coherence, and relevance. Our longitudinal analysis uncovers a structural transformation: pre-2015 coverage emphasized risk and regulatory burden; post-Paris Agreement, discourse shifted toward economic opportunity and innovation, with financial institutions becoming dominant voices. Methodologically, we provide a replicable paradigm for longitudinal media analysis with LLMs; substantively, we reveal how financial elites have internalized and reframed the climate crisis across two decades.
\end{abstract}

\section{Introduction}

Financial news media serves as the nervous system of the global economy, informing not only investors but also shaping corporate strategy and capital allocation decisions worth trillions of dollars \citep{shiller2019}. How this elite domain portrays climate change is therefore consequential, influencing whether climate risks are framed as costs to be minimized or opportunities to be seized. While climate communication in general news has been extensively studied \citep{schmid-petri2017}, the high-stakes arena of financial news remains underexplored, despite its significant role in shaping market expectations and influencing policy debates.

This paper addresses this gap through a longitudinal computational analysis of the \textit{Dow Jones Climate News Corpus}, a collection of 980{,}061 climate-related articles (2000–2023). We focus our extraction on a stratified and uncertainty-enriched sample of 4{,}143 articles that preserves the temporal and thematic diversity of the full dataset. Moving beyond topic models and dictionary-based methods, we introduce an \textbf{Actor–Frame–Argument (AFA)} methodology that leverages Large Language Models (LLMs) to identify who speaks, how climate issues are framed, and which arguments are advanced. To ensure reliability across two decades of discourse, we validate our pipeline against a 2{,}000-article human-annotated gold standard and introduce a \textbf{Decompositional Verification Framework (DVF)} that evaluates extractions along completeness, faithfulness, coherence, and climate relevance with multi-judge validation.

We organize the study around three guiding research questions:
\begin{itemize}
    \item \textbf{RQ1 (Actors):} Which types of financial and policy actors dominate climate discourse in elite financial news, and how has their prominence evolved over time?
    \item \textbf{RQ2 (Frames):} How have climate change frames evolved between 2000--2023, and what external events coincided with major shifts in framing?
    \item \textbf{RQ3 (Arguments):} What argumentative strategies and warrants are deployed to justify climate positions, and how do these differ across actor groups and eras?
\end{itemize}

Our contributions are threefold. First, we develop and validate an LLM-based AFA pipeline for longitudinal extraction of actors, frames, and arguments, anchored by DVF to ensure robustness and auditability. Second, we provide the most comprehensive map to date of climate framing in elite financial news, documenting systematic differences across time, actor groups, and argumentative repertoires. Third, we uncover a structural transformation in financial climate narratives: from a pre-2010 emphasis on \textit{costs and compliance} to a post-Paris Agreement framing of \textit{opportunity and innovation}, linked to external shocks such as the Paris Agreement and subsequent policy shifts.

By combining methodological innovation with substantive insight, this work establishes a replicable paradigm for large-scale discourse analysis.

\section{Related Work}

\subsection{Argument Mining and Discourse Analysis}

Argument mining seeks to recover argumentative structure—claims, evidence, and warrants from text \citep{lawrence-reed-2019-survey}. Early work established end-to-end pipelines for component detection and relation parsing, often with global inference \citep{stab-gurevych-2017}. Subsequent neural approaches improved robustness by modeling non-tree structures and joint decoding of components and links \citep{liu-etal-2021-neural-transition}. Parallel strands in discourse parsing provide complementary representations of coherence that have been leveraged for argument quality and relation modeling \citep{liu-etal-2020-discourse, prasad-etal-2019-pdtb3}. 

Other research has expanded task coverage (stance, evidence retrieval, relation typing, and quality) and domains, utilizing larger and more diverse datasets \citep{hua-etal-2022-iam, chen-etal-2021-marq, gleize-etal-2020-efficient-quality}. Transformer baselines remain strong for span/link prediction and cross-domain transfer, while graph-based architectures and discourse-informed models help capture long-range structure and argumentative coherence \citep{ruiz-dolz-2021-transformer, toledo-etal-2021-graph-quality}. 

Most closely related to our aims are works that examine argument structure at scale and across time. However, existing studies typically operate on short-horizon corpora (essays, forums, debates) and evaluate with intrinsic metrics alone. Emerging evidence shows that properly prompted LLMs can match or outperform task-specific models for argument mining, but concerns remain about faithfulness, bias, and reproducibility \citep{gorur-etal-2025-large, li2025largelanguagemodelsargument}.

\subsection{Media Framing and Climate Communication}

Framing theory emphasizes that how issues are presented strongly conditions public interpretation and policy response \citep{entman1993}. Early studies relied on manual content analysis of small corpora, often in the U.S. or Europe, highlighting political polarization and media biases in the portrayal of climate change \citep{nisbet-2009-frames, schmidpetri2017, boykoff2004}. More recent work has extended this line to multi-country comparisons, showing persistent divergences between U.S., European, and Chinese media in how climate debates are framed \citep{schmidpetri2016, duan_miller_2021_journalism}.

Computational approaches have sought to scale framing analysis. Dictionary-based methods \citep{card-etal-2015} and topic models \citep{blei2003, roberts2014structural} have been applied to climate corpora to identify high-level themes such as risk, adaptation, and responsibility \citep{schirmag2025neural}. However, such approaches often lack granularity and provide limited insight into how actors strategically deploy frames over time. Recent advances in computational framing highlight the need for models that capture not just what frames are invoked but also the argumentative strategies that support them \citep{demszky-etal-2019-analyzing}. Recent work has begun modeling narrative framing in news media, capturing conflicts, resolutions, and entity roles (heroes, victims, villains) in climate discourse \citep{frermann-etal-2023-conflicts, otmakhova-frermann-2025-narrative}, though these approaches focus on general news rather than the specialized financial domain.

\subsection{LLMs for Climate and Financial Discourse}
\label{sec:llm-climate-finance}

Recent work has begun to apply large language models to climate and finance text, yielding methods and findings directly relevant to our Actor-Frame-Argument (AFA) pipeline. For example, \cite{leippold2023thus} explored structured interviews of GPT-3 on climate finance narratives, while \cite{jain2024empowering} demonstrate LLM-based tools for climate-aware investment decision support. Domain-specialized models such as ClimateBERT \citep{webersinke2021climatebert} and SusGen-GPT \citep{masson2024susgengpt} provide useful embeddings and generation priors for climate and sustainability text. Other studies used LLMs to estimate public opinion and targeting effects around climate topics \citep{lee2024can, islam2024posthoc}, and to analyze bias and representational differences in LLM outputs on sociopolitical issues, including climate \citep{cheng2022gpt3}.

\subsection{Our Contributions}

Our work advances these three research strands in complementary ways. Relative to argument mining (§2.1), we target a multi-decade financial news corpus enabling longitudinal analysis, decompose extraction into an integrated Actor–Frame–Argument framework jointly analyzing who speaks, how issues are framed, and which arguments are deployed, and pair LLM extraction with a Decompositional Verification Framework auditing completeness, faithfulness, coherence, and relevance. Relative to framing research (§2.2), we focus on financial news, a high-stakes domain where framing influences capital allocation, and analyze actor-specific rhetorical strategies over time. While narrative framing models capture story-level conflicts and resolutions, our approach emphasizes the economic and regulatory dimensions distinctive to financial discourse. Relative to LLM applications (§2.3), we validate extraction through multi-judge verification against human annotations, emphasizing reproducibility and faithfulness over predictive performance alone.

\section{The Dow Jones Climate News Corpus}

\subsection{Corpus Construction}
We construct the Dow Jones Climate News Corpus in two steps. First, we filter articles in the Dow Jones Newswire (2000–2023) using Dow Jones Intelligent Identifiers (DJIDs), a proprietary subject taxonomy that consistently categorizes financial news \citep{dowjones2024djid,proquest2024factiva}. We retain articles tagged with climate-related DJIDs spanning three domains: (1) \textit{Core Climate Issues} (e.g.,N/CO2 [Carbon Dioxide] ) (2) \textit{Energy Transition} (e.g., N/BFL [Biofuels]), and (3) \textit{Climate-Affected Sectors} (e.g., N/AGR [Agriculture]). A full list of DJIDs and their descriptions is provided in Appendix~\ref{app:djid}.

Second, to quantify potential selection bias introduced by reliance on DJID codes, we conduct a validation study combining keyword-based re-screening and supervised classification. A summary of results indicates that DJID filtering achieves high precision ($\sim$95\%) and recall ($\sim$92\%), suggesting that our climate corpus is both accurate and broadly representative. Full methodological details are provided in Appendix~\ref{app:djid-validation}.

Preprocessing involved standard normalization and near-duplicate removal. Using locality-sensitive hashing, we removed $\sim$8.7\% of articles with high overlap, yielding \textbf{980,061 unique climate-related articles} with full metadata and timestamps for diachronic analysis (Figure~\ref{volume-appendix} in Appendix~\ref{app:volume} shows the temporal distribution of the Dow Jones Climate News Corpus, annotated with major climate policy events). Full preprocessing details are provided in Appendix~\ref{app:preproc}.

\subsection{Sampling Methodology}
\label{app:sampling}

To construct a tractable yet representative subset of our climate corpus, we implemented a four-stage hierarchical sampling procedure. This design ensures that the 4,143-article sample retains the temporal and thematic diversity of the full 980k-article corpus, while also enriching for complex argumentative texts.

\subsubsection{Temporal Stratification}
We stratified the corpus into four periods reflecting major phases in climate finance discourse: Pre-crisis (2000–2007), Financial Crisis (2008–2012), Post-crisis Recovery (2013–2018), and Climate Finance Surge (2019–2023). Within each stratum $t$, the sample size was determined by proportional allocation:
\[
n_t = \left\lfloor n_{\text{total}} \times \frac{|C_t|}{|C|}\right\rfloor ,
\]
where $|C_t|$ is the size of stratum $t$ and $|C|$ is the size of the full corpus.

\subsubsection{Thematic Clustering}
To preserve thematic coverage, each article was represented by a 2:1 weighted concatenation of its headline and first paragraph embeddings, computed using the SBERT \texttt{all-MiniLM-L6-v2} model \cite{reimers-2019-sentence-bert}. We applied agglomerative clustering with Ward linkage \cite{ward-1963-hierarchical}, selecting the optimal number of clusters $k_t$ by maximizing the silhouette score \cite{rousseeuw-1987-silhouettes} within the range $[5, \min(15, \lfloor n_t/20 \rfloor)]$. Pilot studies indicated that $k<5$ conflated distinct themes, while $k>15$ produced redundant micro-clusters.

\subsubsection{Representative Article Selection}
From each cluster, we applied Maximal Marginal Relevance (MMR) \cite{carbonell-1998-mmr} to select representative articles that balance centrality and diversity:
\[
\text{MMR}(a_i) = \lambda \cdot \text{sim}_c(a_i) - (1-\lambda) \cdot \max_{a_j \in S} \text{sim}_s(a_i, a_j).
\]
We set $\lambda=0.7$, slightly favoring centrality. A sensitivity analysis across $\lambda \in [0.5, 0.9]$ showed stable downstream performance (Appendix~\ref{app:sample}).

\subsubsection{Active Learning Enrichment}
To enrich the sample with harder argumentative cases, we incorporated an active learning loop. A preliminary RoBERTa-base argument detector (F1 = 79.5\% on a seed set) was applied to the wider corpus (Appendix~\ref{app:AL}). Articles with prediction entropy above the 90th percentile were prioritized. Iterative sampling converged after four rounds, measured by Jensen–Shannon divergence of entropy distributions ($<$0.05 between iterations). 

\subsubsection{Final Sample}
The resulting 4,143-article sample achieves high fidelity to the original corpus: Jensen–Shannon divergence $<$0.1 across temporal and thematic distributions, cosine similarity $>$0.85 between sample and population centroids, and minimal degradation ($-3.3$ points) in downstream argument extraction performance compared to a 20k random sample.

\begin{table}[t]
\centering
\small
\begin{tabular}{lrrr}
\toprule
\textbf{Temporal Stratum} & \textbf{Original} & \textbf{Sample} & \textbf{\%} \\
\midrule
Pre-crisis (2000--2007) & 5,441 & 155 & 0.8 \\
Financial crisis (2008--2012) & 320,157 & 1,355 & 0.8 \\
Post-crisis (2013--2018) & 215,351 & 911 & 0.8 \\
Climate surge (2019--2023) & 439,112 & 1,722 & 0.8 \\
\midrule
\textbf{Total} & \textbf{980,061} & \textbf{4,143} & \textbf{0.8} \\
\bottomrule
\end{tabular}
\caption{Distribution of articles across temporal strata in original corpus and strategic sample.}
\label{tab:sample_distribution}
\end{table}

\section{Actor-Frame-Argument (AFA) Extraction Pipeline}

\label{sec:pipeline}

\begin{figure*}[t]
    \centering
    \includegraphics[width=\textwidth]{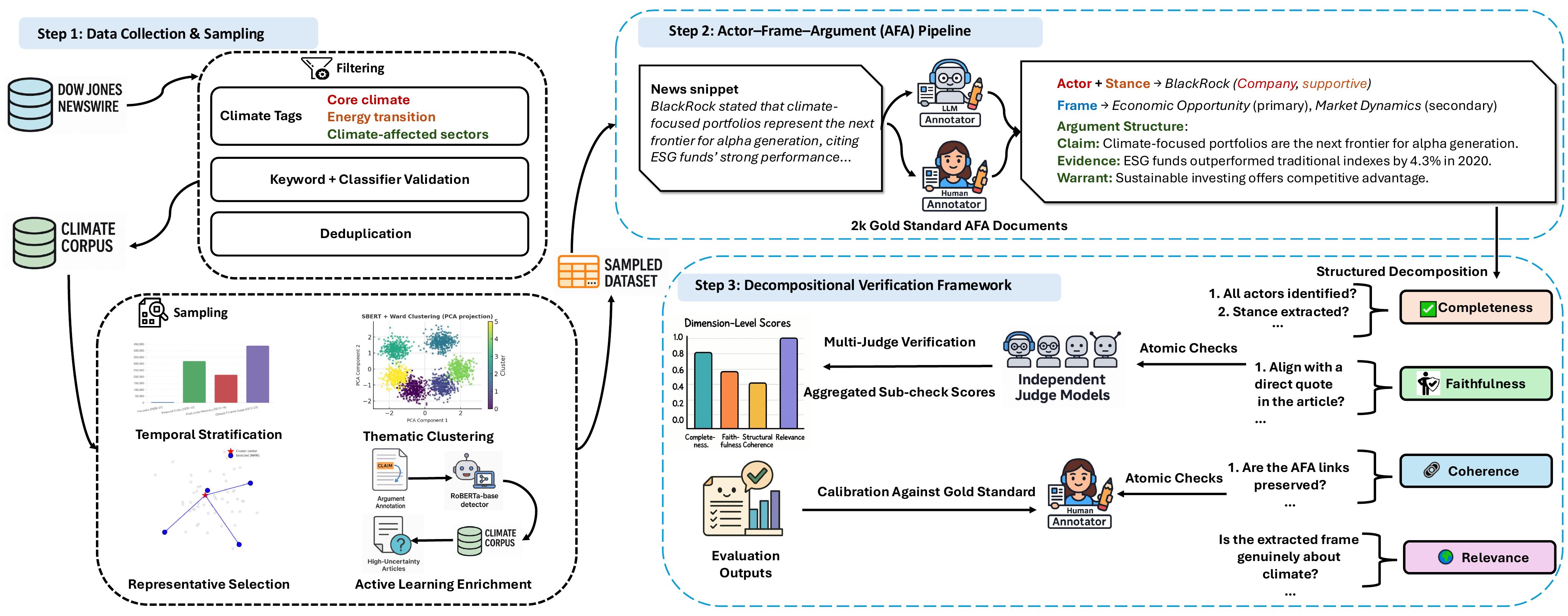}
    \caption{End-to-end pipeline of our study. 
    Step 1: Data collection and sampling from the Dow Jones Climate Corpus, including stratification, clustering, and active learning enrichment. 
    Step 2: Actor–Frame–Argument (AFA) extraction pipeline, identifying actors, frames, and argument structures. 
    Step 3: Decompositional verification framework, combining multi-judge evaluation with validation against a human-annotated gold standard.}
    \label{fig:pipeline}
\end{figure*}

We design a modular pipeline for extracting Actor–Frame–Argument (AFA) structures from financial news. The goal is to represent who speaks, how climate change is framed, and what argumentative strategies are deployed, in a form that is scalable and reproducible. The pipeline is sequential; each stage conditions on the previous one, ensuring coherence across actors, frames, and claims. Figure \ref{fig:pipeline}illustrates the pipeline architecture.

\paragraph{Actor–Stance Identification.} 
The first stage identifies actors and their expressed stances on climate issues. Each extraction specifies the actor \textit{name}, \textit{type} 
(\textit{company}, \textit{financial institution}, \textit{government}, \textit{NGO/advocacy}, \textit{individual}), and expressed \textit{stance}.
, and \textit{stance} (\textit{supportive}, \textit{opposing}, \textit{neutral}, or \textit{mixed}). For attributional fidelity, every extraction is linked to a direct \textit{evidence} quote from the article.

\paragraph{Frame Classification.}
The second stage assigns one \textit{primary frame} and an optional \textit{secondary frame} from our eight-frame typology.

The eight-frame typology adapts established frameworks from climate communication research \citep{nisbet-2009-frames, schmid-petri-2017-climate} to financial discourse. Key adaptations include: (1) splitting generic ``Economic Development'' into \textit{Economic Risk} vs.\ \textit{Economic Opportunity} to capture the critical distinction in financial contexts, (2) adding \textit{Market Dynamics} to capture competitive positioning unique to financial news, and (3) refining ``Morality/Ethics'' to \textit{Social Responsibility} reflecting corporate framing conventions. This typology emerged through pilot coding of 500 articles. Full mapping to source frameworks is in Appendix~\ref{app:frames}.

\paragraph{Argument Extraction.} 
Finally, the model decomposes the article’s argumentation into its components: a central \textit{claim}, supporting \textit{evidence}, and a \textit{warrant} connecting claim and evidence. 

\subsection{Models and Prompting}

While domain-specific encoders such as ClimateBERT \citep{webersinke2021climatebert} or data-centric models like SusGen-GPT \citep{masson2024susgengpt} offer attractive pretraining priors for climate text, our focus is on extraction robustness and longitudinal generalization across heterogeneous financial-news styles.

We therefore adopt a closed–open pairing strategy for extraction. The primary extractor is Gemini-2.5-flash \cite{comanici2025gemini25}, which offers strong instruction-following reliability and efficiency. To ensure that findings are not from a single proprietary model, we benchmark the same prompts on a diverse open-weight model LLaMA-4 Maverick-17B \cite{meta2025llama4}. This pairing balances performance (closed-sourced model) with reproducibility and transparency (open-sourced model), following best practices in recent work on LLM-as-annotator pipelines \cite{tan-etal-2024-large, tornberg2024best, pavlovic-poesio-2024-effectiveness}. 

While multiple open-weight models exist, we focus on one representative baseline in the main analysis for clarity.  All prompts use a structured chain-of-thought design and are released Appendix~\ref {app:prompts} for reproducibility. 

\subsection{Human-Annotated Gold Standard}
To establish a reliable evaluation anchor, we constructed a 2,000-article gold-standard set drawn from our corpus. Each annotation unit is a complete article (mean=1,087 tokens, median=982 tokens). The ``news snippet'' in Figure~\ref{fig:pipeline} is illustrative; actual processing uses full articles. Articles were annotated on the Zooniverse citizen-science platform (see Appendix~\ref{app:gold}), with each item receiving at least five independent annotations. For categorical tasks (stance, frame), final labels were obtained by majority vote. For span-based tasks (actors, arguments), we applied a two-step consensus procedure: token-level agreement scores were computed across annotators, tokens with majority support were retained, and contiguous majority-agreed tokens were merged into unified spans (see Appendix ~\ref{app:span_consensus}). This aggregation preserves boundary precision for short mentions while preventing over-fragmentation in longer argument spans.

We computed inter-annotator agreement (IAA) using metrics appropriate to multi-annotator settings: pairwise $F_1$ for actor spans \citep{stefan2017argument}, Krippendorff’s~$\alpha$ \citep{krippendorff2004content} for stance, Krippendorff’s~$\alpha$ for frames, and span-level $F_1$ for argument claim extraction \citep{stab2014annotating,habernal2018argument}. Table~\ref{tab:iaa} summarizes observed IAA scores, all of which fall in the “substantial agreement” range ( > 0.7) \cite{landis1977measurement}, confirming the reliability of the citizen-science annotation protocol.

\begin{table}[t]
\centering
\footnotesize
\begin{tabularx}{\columnwidth}{l c c}
\toprule
\textbf{Component} & \textbf{Metric} & \textbf{IAA} \\
\midrule
Actor Identification & Pairwise F1 & 0.81 \\
Stance Classification & Krippendorff’s~$\alpha$ & 0.76 \\
Frame Assignment & Krippendorff’s $\alpha$ & 0.72 \\
Claim Extraction & Span-level F1 (macro) & 0.74 \\
\bottomrule
\end{tabularx}
\caption{Inter-annotator agreement across AFA components under the Zooniverse citizen-science protocol.}
\label{tab:iaa}
\end{table}

\subsection{Decompositional Verification Framework}

To ensure reliable and auditable evaluation, we introduce a 
\textbf{Decompositional Verification Framework (DVF)} that 
pushes beyond single holistic scores. Instead of asking judges to rate an extraction globally, DVF decomposes the evaluation into 
fine-grained sub-checks, verifies them across multiple model 
families, and anchors them to human gold annotations.

\paragraph{Structured Decomposition.} 
DVF evaluates extractions along four dimensions: 
(1) \textit{Completeness} (all components captured), 
(2) \textit{Faithfulness} (alignment with source text), 
(3) \textit{Coherence} (schema and actor–frame–argument links), 
and (4) \textit{Relevance} (domain specificity). 
Full sub-checks are provided in Appendix~\ref{app:dvf}.

\paragraph{Multi-Judge Verification.} 
To mitigate self-grading bias, DVF employs a diverse set of 
judges: GPT-4o \cite{openai2024gpt4o}, Claude-Sonnet-4 \cite{anthropic2024claudesonnet4}, and two open-weight evaluators (Qwen3-30B A3B \cite{qwen3_30b_a3b}, Mixtral-8$\times$22B \cite{mixtral_8x22b}). 
Judges provide sub-check ratings, which are aggregated into dimension-level scores. To anchor these automated scores, we created a human evaluation set of 500 randomly sampled LLM outputs. These outputs were annotated under the DVF rubric by coders distinct from the AFA gold-standard annotators. This separate evaluation set enables validation of automated DVF scores, ensuring that judge reliability is grounded in human assessments while avoiding circularity.

\section{Results and Discussion}
\label{sec:results}

We present our findings in four parts. First, we validate the reliability of the AFA extraction pipeline (§ \ref{subsec:validation}). Second, we analyze longitudinal trends in actor prominence using our five-category schema (§ \ref{subsec:actors}, RQ1). Third, we document the structural transformation in framing strategies across two decades (§ \ref{subsec:transformation}, RQ2). Finally, we examine how actors combine frames with argumentative strategies, revealing distinct rhetorical profiles (§ \ref{subsec:alignment}, RQ3). A qualitative error analysis of 150 sampled mispredictions is provided in Appendix~\ref{app:error-analysis}.

\subsection{Pipeline Validation}
\label{subsec:validation}
\subsubsection{Justification for LLM-Based Extraction}
\label{sec:llm-justification}

A critical question is whether LLMs' computational cost is justified when fine-tuned Pre-trained Language Models (PLMs) could achieve similar performance on specific tasks. We test two hypotheses: (H1) \textit{LLMs provide superior performance on complex, multi-class tasks }, and (H2) \textit{sequential dependencies through in-context learning will improve the performance}.

\paragraph{Experimental Setup}
We partitioned our 2,000-article gold standard into 1,400 train / 300 validation / 300 test splits. We compared three RoBERTa-Large \cite{zhuang-etal-2021-robustly} baselines against our zero-shot LLM pipeline:

\textbf{Independent RoBERTa:} Each AFA component trained as an isolated model receiving only raw article text: (1) token-level NER for actors, (2) sequence classifier for stance, (3) multi-label classifier for frames, and (4) span extractor for arguments.

\textbf{Sequential RoBERTa (Gold):} Downstream components receive gold-standard upstream annotations as additional features (e.g., frame classifier receives gold actors concatenated with article text).

All RoBERTa models used identical hyperparameters. Training details are provided in Appendix K.

\paragraph{Results}
Table~\ref{tab:llm-justification} shows LLMs outperform fine-tuned RoBERTa-Large across all components, with largest gaps on frame classification ($+15.2$ F1) and argument extraction ($+17.2$ F1), which supports H1. 

For example, RoBERTa classifies ``renewable energy fund outperformed traditional portfolios by 4.3\%'' as \textit{Technological Solution} (triggered by ``renewable energy''), while the correct frame is \textit{Economic Opportunity} (centered on financial returns). LLMs correctly identify the argumentative focus through longer-range contextual reasoning. 

RoBERTa also struggles with long-range dependencies in complex argumentative structures. We found out that it either truncates claims prematurely at clause boundaries or over-extends into supporting evidence, suggesting insufficient training data for learning nuanced argument boundaries.

\begin{table}[t]
\centering
\small
\begin{tabular}{@{}lcccc@{}}
\toprule
\textbf{Model} & \textbf{Actor} & \textbf{Stance} & \textbf{Frame} & \textbf{Arg.} \\
\midrule
Independent RoBERTa & .762 & .718 & .504 & .563 \\
+ Gold upstream & .762 & .742 & .631 & .595 \\
\midrule
\textbf{Gemini-2.5} & \textbf{.819} & \textbf{.791} & \textbf{.783} & \textbf{.767} \\
\midrule
$\Delta$ (LLM -- Best RoBERTa) & +5.7 & +4.9 & +15.2 & +17.2 \\
\bottomrule
\end{tabular}
\caption{Performance comparison (macro F1) on 300-article test set.}
\label{tab:llm-justification}
\end{table}

\paragraph{Sequential Dependencies and Error Propagation}
Table~\ref{tab:llm-justification} reveals the critical finding supporting H2: \textit{Sequential conditioning is beneficial}: Gold upstream features improve RoBERTa performance substantially ($+12.7$ for frames, $+3.2$ for arguments), confirming that AFA components are not independent, frames depend on actors, and arguments depend on both. But independent RoBERTa classifiers can not naturally achieve that.

\paragraph{Implications}
To summarize, for our use case (multi-decade, multi-component extraction with limited human annotations), LLMs provide superior cost-performance trade-offs. They excel at: (1) complex multi-class tasks requiring semantic distinction and (2) sequential reasoning through in-context learning rather than brittle feature engineering. 

\subsubsection{LLM-based Performance}
To establish the reliability of our LLM-based extraction pipeline, we evaluate model outputs against a 2,000-article gold standard created via Zooniverse annotations. 

Table~\ref{tab:pipeline_performance} reports system performance across all three AFA components for both Gemini-2.5-flash (primary extractor) and LLaMA-4-Maverick-17B (open-weight baseline). Gemini consistently outperforms LLaMA by roughly two to three points across components. Performance is highest for actor type identification ($F_1=0.847$) and lowest for second frame classification ($F_1=0.692$), reflecting the inherent ambiguity and sparsity of secondary frame labels. Unlike primary frames, which typically correspond to the dominant narrative focus of a passage, secondary frames often capture peripheral or overlapping interpretive cues (e.g., regulatory and technological frames co-occurring), making automatic prediction more difficult.

Beyond intrinsic F1, we validate extractions using the DVF aggregate scores, which confirm quality across four dimensions. Faithfulness in particular remains consistently high, indicating that extracted components are well-grounded in the source text. To ensure transparency, DVF scores are validated against a 500-sample human evaluation set. Detailed per-judge breakdowns and human validation statistics are reported in Appendix~\ref{app:dvf}, while we present only the aggregate results in the main text for clarity. 

\begin{table}[t]
\centering
\small
\begin{tabular}{lcc}
\toprule
\textbf{Component} & \textbf{Gemini-2.5} & \textbf{LLaMA-4} \\
\midrule
\multicolumn{3}{l}{\textit{Actor--Stance Identification}} \\
\quad Actor Type & \textbf{0.847} & 0.823 \\
\quad Stance & \textbf{0.791} & 0.768 \\
\quad Overall F1 & \textbf{0.819} & 0.796 \\
\midrule
\multicolumn{3}{l}{\textit{Frame Classification}} \\
\quad Primary Frame & \textbf{0.783} & 0.761 \\
\quad Secondary Frame & \textbf{0.692} & 0.671 \\
\midrule
\multicolumn{3}{l}{\textit{Argument Extraction}} \\
\quad Claim & \textbf{0.806} & 0.782 \\
\quad Evidence & \textbf{0.774} & 0.753 \\
\quad Warrant & \textbf{0.721} & 0.698 \\
\quad Overall F1 & \textbf{0.767} & 0.744 \\
\midrule
\multicolumn{3}{l}{\textit{DVF Aggregate Scores}} \\
\quad Completeness & \textbf{0.831} & 0.809 \\
\quad Faithfulness & \textbf{0.887} & 0.863 \\
\quad Coherence & \textbf{0.792} & 0.771 \\
\quad Relevance & \textbf{0.856} & 0.834 \\
\bottomrule
\end{tabular}
\caption{Pipeline performance against the 2{,}000-article gold standard (macro F1). 
Gemini-2.5-flash serves as the primary extractor; LLaMA-4-Maverick-17B provides open-weight baseline performance. 
Bolded values indicate the best performance per component.}
\label{tab:pipeline_performance}
\end{table}

\subsection{Actor Prominence Over Time}
\label{subsec:actors}

\begin{figure}[t]
\centering
\includegraphics[width=\linewidth]{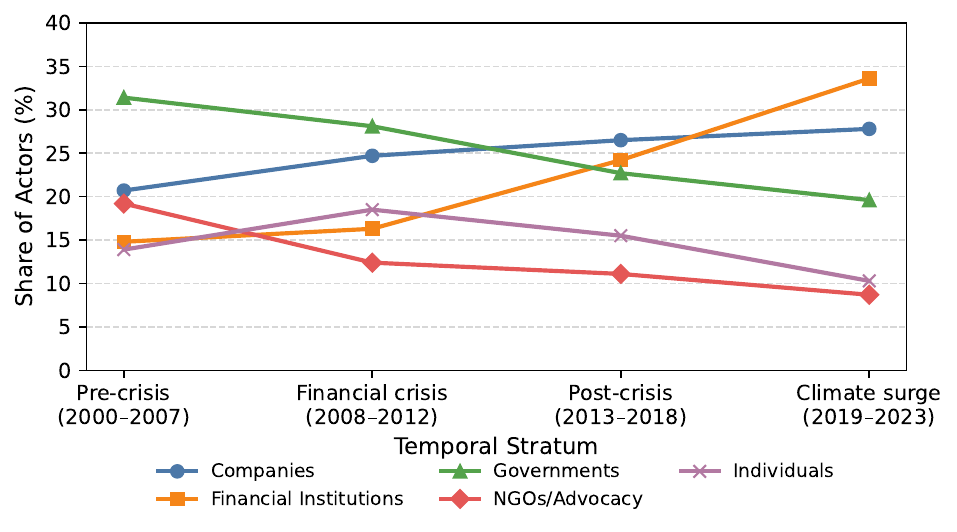}
\caption{Actor prominence over time, showing proportional distribution across five actor categories in four temporal strata.}
\label{fig:actor_trends}
\end{figure}

To address \textbf{RQ1 (Actors)}, we examine how different categories of actors participate in climate discourse in financial news. Using our five-category schema 
(\textit{companies}, \textit{financial institutions}, \textit{governments}, 
\textit{NGOs/advocacy}, \textit{individuals}), we track longitudinal changes in actor prominence across four temporal strata (2000--2007, 2008--2012, 2013--2018, 2019--2023). Figure~\ref{fig:actor_trends} displays actor distributions over time.  

We observe a clear structural shift in discursive authority
($\chi^2 = 217.3$, $p < 0.001$). In the early 2000s,
government ($\approx 31\%$) and companies ($\approx 21\%$) dominated climate-related financial reporting, consistent with a regulatory- and advocacy-driven framing of climate change. For instance, a 2006 article cites the World Wildlife Fund warning that ``regulatory inaction risks locking in high-carbon infrastructure.'' During the financial crisis, individuals rose to $\approx 18\%$ of mentions, while governments began to decline in relative presence ($\approx 28\%$), reflecting a shift toward market-based voices.

In the post-crisis period, coinciding with the Paris Agreement,
financial institutions surged from $\approx 15\%$ before the crisis to more than $25\%$ after 2015, and grew to $\approx 34\%$ during recent years, becoming the single largest actor group. Asset managers and banks increasingly positioned themselves as climate leaders. For example, a 2021 Dow Jones piece quotes BlackRock: ``Sustainable finance is now central to alpha generation.''

In parallel, during Climate surge NGOs fell below 10\%, while governments stabilized around 20\%, often providing background regulatory context rather than leading the narrative.  

These findings suggest a rebalancing of discursive power: from a regime led by governments to a financial-market regime where financial institutions and companies are the dominant voices in elite climate discourse. Exact proportions are reported in Appendix~\ref{app:actor_table}.

\subsection{Narrative Transformation in Frames}
\label{subsec:transformation}

To address \textbf{RQ2 (Frames)}, we analyze temporal shifts in how climate issues are framed in elite financial news. We track the eight-frame typology introduced earlier (economic risk, regulatory compliance, economic opportunity, technological solution, market dynamics, environmental urgency, social responsibility, uncertainty/skepticism) across four temporal strata, and test for significant distributional changes over time.

\begin{figure}[t]
\centering
\includegraphics[width=\columnwidth]{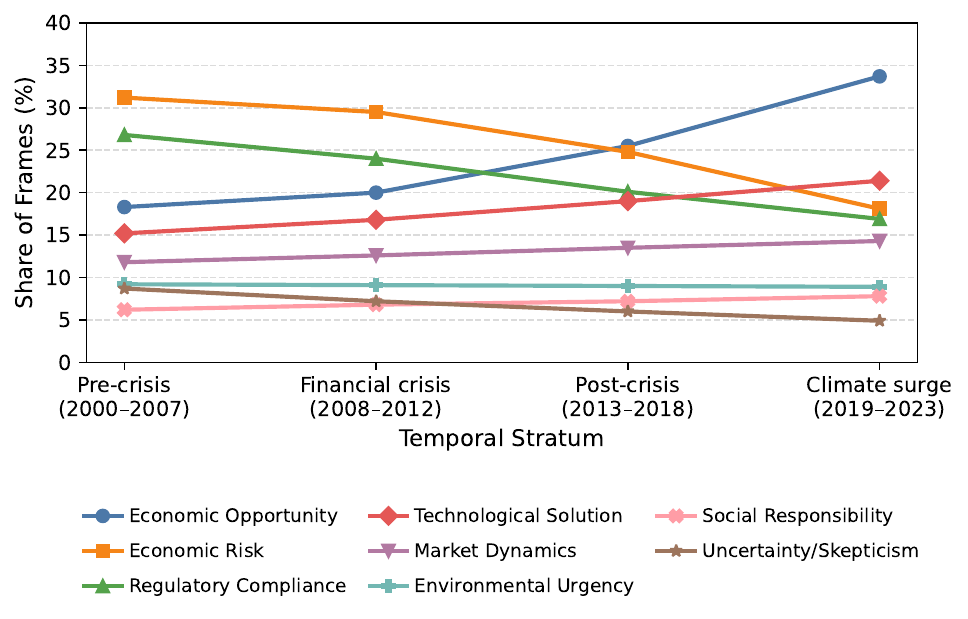}
\vspace{-1mm}
\caption{Temporal dynamics of climate frames across four strata.}
\vspace{-2mm}
\label{fig:frame_trends}
\end{figure}

\paragraph{From risk/compliance to opportunity/innovation.}
 A visual overview in Figure~\ref{fig:frame_trends} shows a clear trend: risk and compliance-oriented coverage recedes while opportunity and innovation-oriented coverage grows. A changepoint analysis (PELT) identifies a statistically significant structural break in \textbf{2015Q4--2016Q2}, aligning with the Paris Agreement and related policy announcements, after which \textit{economic opportunity} and \textit{technological solution} become increasingly salient.

\paragraph{Statistical evidence.}
We observe a reversal in the relative weights of \textit{economic risk} and \textit{economic opportunity} frames before vs.\ after the 2015–2016 changepoint identified by our PELT analysis (see §5.3). Opportunity rises substantially in the late period, while risk declines; \textit{technological solution} also increases and frequently co-occurs with opportunity (positive association), consistent with a forward-looking investment narrative. Exact per-period percentages and significance tests are reported in Appendix~\ref{app:frame_table}.

\paragraph{Illustrative contrast.}
Early-period articles often emphasize compliance costs or exposure (e.g., mandated adjustments, liabilities), whereas post-2015 coverage increasingly highlights growth and competitive positioning (e.g., green finance pipelines, innovation leadership). Short examples are shown below (paraphrased for brevity):
\begin{itemize}\itemsep0.2em
  \item \textbf{Risk (pre-2015):} “New carbon rules will raise operating costs and pressure margins.”
  \item \textbf{Opportunity (post-2015):} “Climate-focused portfolios and clean-tech investments open new alpha and market-leadership avenues.”
\end{itemize}

There is a clear, statistically grounded pivot in financial-news framing of climate: the narrative transitions from \textit{risk and regulatory burden} to \textit{opportunity and technological innovation}, with the turning point concentrated around 2015–2016. This reframing is consistent with the sector’s shift toward green finance and investment-led rationales.

\vspace{-2mm}

\subsection{Actor–Frame–Argument Alignment (RQ3)}
\label{subsec:alignment}

To answer \textbf{RQ3 (Arguments)}, we analyze how actor groups combine frames with arguments to construct climate narratives. Beyond overall shifts, this analysis reveals distinct rhetorical repertoires that different actors use in elite financial news.

\paragraph{Actor--Frame Associations.}
Table~\ref{tab:actor_frame_matrix} reports standardized residuals from a $\chi^{2}$ test of independence, capturing which frames are over- or under-represented by each actor category. Companies and financial institutions show a pronounced 
preference for \textit{economic opportunity} frames, while NGOs disproportionately emphasize \textit{environmental urgency}. Governments and regulators display more balanced use of \textit{risk} and \textit{regulatory compliance} frames, reflecting their institutional roles.

\paragraph{Argument Strategies by Actor.}
Warrant analysis highlights divergent argumentative logics:
\begin{itemize}\itemsep0.2em
  \item \textbf{Companies} often invoke \textit{competitive advantage} and 
  \textit{innovation opportunity} warrants, presenting climate action as a business growth strategy.
  \item \textbf{Financial Institutions} emphasize \textit{market leadership} and \textit{risk-adjusted returns}, linking climate investments to fiduciary responsibility.
  \item \textbf{Governments/Regulators} rely heavily on \textit{regulatory necessity} and \textit{compliance} warrants.
  \item \textbf{NGOs/Advocacy groups} foreground \textit{environmental urgency} and 
  \textit{ethical responsibility}.
  \item \textbf{Individuals (researchers/experts)} emphasize \textit{scientific evidence} and risk communication, often reinforcing NGO frames but with technical grounding.
\end{itemize}

\paragraph{Argument Complexity.}
We measure elaboration as the median number of premises per claim. Opportunity-oriented arguments show greater complexity (median 2.8–3.2 premises), while risk arguments are terser (median 1.4–1.7 premises), which suggests that forward-looking frames demand more justificatory investment.

\paragraph{Illustrative Examples.}
\begin{itemize}\itemsep0.3em
  \item \textbf{NGO (2010)}: “Governments must act now; unchecked emissions will accelerate ecological collapse.” (Frame: Environmental Urgency).
  \item \textbf{Company (2018)}: “Investing in clean tech secures long-term competitiveness in global markets.” (Frame: Economic Opportunity).
  \item \textbf{Financial Institution (2021)}: “Climate-focused portfolios represent the next frontier for alpha generation.” (Frame: Economic Opportunity).
\end{itemize}

Actor groups not only favor different frames but also use distinctive argumentative strategies. This alignment of actors, frames, and arguments demonstrates how the narrative has shifted: financial institutions and companies increasingly position themselves as market leaders in the climate transition, while NGOs emphasize urgency and governments focus on compliance. Together, these patterns reveal a fragmented but structured rhetorical field in financial climate discourse.

These financial-news findings align with LLM-based studies of climate discourse outside the financial domain. For example, Lee et al. \citep{lee2024can} show that model-derived framing signals correlate with public opinion trends on global warming, suggesting that automated frame extraction can reflect wider societal narratives. This cross-domain concordance lends external validity to our inferences about the reframing of climate in financial media.

\section{Conclusion}
\label{sec:conclusion}

We presented an Actor–Frame–Argument (AFA) pipeline for longitudinal discourse analysis that combines LLM-based extraction with a decompositional verification framework. Applied to a corpus of climate-related financial news spanning two decades, the approach yields both methodological robustness and substantive insight. 

We document a structural transformation in elite financial coverage: from early emphasis on economic risk and regulatory compliance toward a post-2015 discourse centering on opportunity and technological solutions. We further show that actors deploy distinct rhetorical strategies, companies and financial institutions emphasize competitive advantage and risk-adjusted returns, governments stress compliance and policy instruments, and NGOs foreground environmental urgency, revealing a patterned but heterogeneous field of climate finance narratives.

Methodologically, our pipeline offers a replicable pipeline for integrating actor identification, framing, and argument mining with auditable evaluation. The results underscore how financial media can shape expectations about the pace and direction of the climate transition.

By bridging advances in NLP with questions central to climate communication and finance, this study shows how computational social science can illuminate the stories that steer capital and policy.

\section*{Limitations}
Our study is based on a single, English-language news source (Dow Jones Newswire). Future work should diversify sources to assess the generality of our findings. While we validated our LLM-based annotation, LLMs have inherent biases. Our validation mitigates this, but cannot eliminate it entirely. Recent work highlights that LLMs can reproduce or amplify biases in climate and sociopolitical communication \citep{cheng2022gpt3, islam2024posthoc}. Our Decompositional Verification Framework reduces some risks by multi-judge validation and faithfulness checks, but a comprehensive fairness audit and subgroup robustness tests remain important future directions, especially for applications that could influence investment or policy decisions. Finally, our study maps discourse but does not directly measure its real-world impact on capital flows, a critical avenue for future research. Future work should extend coverage beyond a single newswire and language, link discursive shifts to measurable market outcomes and policy cycles and explore real-time monitoring tools for emerging narrative turns.

\section*{Ethics Statement}
 Our analysis does not involve private or personally identifiable information. We acknowledge the environmental costs of large model training and inference. Recent analyses indicate that improvements in hardware efficiency and software optimizations have begun to reduce marginal carbon intensity for many workloads \citep{patterson2022carbon}, but these costs remain nontrivial. To reduce our footprint, we used stratified sampling, efficient prompting, and multi-stage extraction rather than retraining large models end-to-end. We report these operational choices to enable reproducibility and to encourage energy-aware replication. Future work should additionally explore lightweight domain adapters and retrieval-augmented approaches to further minimize compute and emissions. We also acknowledge that LLMs may reflect biases from their training data; our validation against human annotators is a crucial check. The methods presented could be used to monitor "greenwashing," but could also be used to craft more persuasive disinformation. We release our findings in the spirit of open academic inquiry, believing the benefit of transparently understanding media narratives outweighs the risk of misuse.

 \section*{Acknowledgments}

Ruiran Su is supported by a Jardine Foundation Scholarship. We gratefully acknowledge the contributions of the volunteers on the Zooniverse platform, whose dedication made the construction of the gold-standard dataset possible. We also thank Dow Jones for providing access to the news corpus used in this study. We also acknowledge the use of the University of Oxford's Advanced Research Computing (ARC) facility for the computational resources used in this work. Finally, we thank the anonymous reviewers for their constructive feedback that helped improve this paper.

\bibliography{emnlp2023-latex/custom}

\begin{thebibliography}{57}
\expandafter\ifx\csname natexlab\endcsname\relax\def\natexlab#1{#1}\fi

\bibitem[{Anthropic(2024)}]{anthropic2024claudesonnet4}
Anthropic. 2024.
\newblock Claude sonnet 4.
\newblock Model documentation / system card.
\newblock Claude Sonnet 4, proprietary model.

\bibitem[{Blei et~al.(2003)Blei, Ng, and Jordan}]{blei2003}
David~M. Blei, Andrew~Y. Ng, and Michael~I. Jordan. 2003.
\newblock \href {https://www.jmlr.org/papers/v3/blei03a.html} {Latent dirichlet allocation}.
\newblock \emph{Journal of Machine Learning Research}, 3:993--1022.

\bibitem[{Boykoff and Boykoff(2004)}]{boykoff2004}
Maxwell~T. Boykoff and Jules~M. Boykoff. 2004.
\newblock \href {https://doi.org/10.1016/j.gloenvcha.2003.10.001} {Balance as bias: Global warming and the u.s. prestige press}.
\newblock \emph{Global Environmental Change}, 14(2):125--136.

\bibitem[{Carbonell and Goldstein(1998)}]{carbonell-1998-mmr}
Jaime Carbonell and Jade Goldstein. 1998.
\newblock \href {https://doi.org/10.1145/290941.291025} {The use of mmr, diversity-based reranking for reordering documents and producing summaries}.
\newblock In \emph{Proceedings of the 21st Annual International ACM SIGIR Conference on Research and Development in Information Retrieval}, pages 335--336, Melbourne, Australia.

\bibitem[{Card et~al.(2015)Card, Boydstun, Gross, Resnik, and Smith}]{card-etal-2015}
Dallas Card, Amber~E. Boydstun, Justin~H. Gross, Philip Resnik, and Noah~A. Smith. 2015.
\newblock \href {https://doi.org/10.3115/v1/P15-2072} {The media frames corpus: Annotations of frames across issues}.
\newblock In \emph{Proceedings of the 53rd Annual Meeting of the Association for Computational Linguistics (ACL) \& 7th International Joint Conference on Natural Language Processing (IJCNLP)}, pages 438--444. Association for Computational Linguistics.

\bibitem[{Chen et~al.(2023)Chen, Duan, Peng, and Zhang}]{cheng2022gpt3}
Kaiping Chen, Ruitong Duan, Ying Peng, and Jingwen Zhang. 2023.
\newblock \href {https://arxiv.org/abs/2209.13627} {How gpt-3 responds to different publics on climate change and black lives matter: A critical appraisal of equity in conversational ai}.
\newblock \emph{arXiv preprint arXiv:2209.13627}.
\newblock Preprint, last revised March 2023.

\bibitem[{Chen et~al.(2021)Chen, He, and Zhang}]{chen-etal-2021-marq}
Zheng Chen, Yulan He, and Yu~Zhang. 2021.
\newblock \href {https://doi.org/10.18653/v1/2021.emnlp-main.558} {Hitting your marq: Multimodal argument quality assessment}.
\newblock In \emph{Proceedings of the 2021 Conference on Empirical Methods in Natural Language Processing (EMNLP)}, pages 6958--6970. Association for Computational Linguistics.

\bibitem[{Comanici et~al.(2025)Comanici, Bieber, Schaekermann, Pasupat, Sachdeva, Dhillon, and et~al.}]{comanici2025gemini25}
Gheorghe Comanici, Eric Bieber, Mike Schaekermann, Ice Pasupat, Noveen Sachdeva, Inderjit Dhillon, and Marcel~Blistein et~al. 2025.
\newblock Gemini 2.5: Pushing the frontier with advanced reasoning, multimodality, long context, and next generation agentic capabilities.
\newblock Technical report, Google / DeepMind.
\newblock Preprint / technical report, includes the Gemini 2.5 Flash model.

\bibitem[{Demszky et~al.(2019)Demszky, Garg, Voigt, Zou, Shapiro, Gentzkow, and Jurafsky}]{demszky-etal-2019-analyzing}
Dorottya Demszky, Nikhil Garg, Rob Voigt, James Zou, Jesse Shapiro, Matthew Gentzkow, and Dan Jurafsky. 2019.
\newblock \href {https://doi.org/10.18653/v1/N19-1304} {Analyzing polarization in social media: Method and application to tweets on 21 mass shootings}.
\newblock In \emph{Proceedings of the 2019 Conference of the North {A}merican Chapter of the Association for Computational Linguistics: Human Language Technologies, Volume 1 (Long and Short Papers)}, pages 2970--3005, Minneapolis, Minnesota. Association for Computational Linguistics.

\bibitem[{{Dow Jones}(2024)}]{dowjones2024djid}
{Dow Jones}. 2024.
\newblock \href {https://developer.dowjones.com/documents/site-docs-factiva_apis-factiva_workflow_apis_rest-factiva_news_search-factiva_djid_taxonomy_api} {{Dow Jones Intelligent Identifiers (DJID) Taxonomy API}}.
\newblock Proprietary subject classification system for financial news content.

\bibitem[{Duan and Miller(2021)}]{duan_miller_2021_journalism}
Ran Duan and Serena Miller. 2021.
\newblock \href {https://doi.org/10.1177/1464884919873173} {Climate change in china: A study of news diversity in party-sponsored and market-oriented newspapers}.
\newblock \emph{Journalism}, 22(10):2493--2510.

\bibitem[{Entman(1993)}]{entman1993}
Robert~M. Entman. 1993.
\newblock \href {https://doi.org/10.1111/j.1460-2466.1993.tb01304.x} {Framing: Toward clarification of a fractured paradigm}.
\newblock \emph{Journal of Communication}, 43(4):51--58.

\bibitem[{Frermann et~al.(2023)Frermann, Li, Khanehzar, and Mikolajczak}]{frermann-etal-2023-conflicts}
Lea Frermann, Jiatong Li, Shima Khanehzar, and Gosia Mikolajczak. 2023.
\newblock \href {https://doi.org/10.18653/v1/2023.acl-long.486} {Conflicts, villains, resolutions: Towards models of narrative media framing}.
\newblock In \emph{Proceedings of the 61st Annual Meeting of the Association for Computational Linguistics (Volume 1: Long Papers)}, pages 8712--8732, Toronto, Canada. Association for Computational Linguistics.

\bibitem[{Gleize et~al.(2020)Gleize, Shnarch, Levy, Bogin, Aharonov, and Slonim}]{gleize-etal-2020-efficient-quality}
Martin Gleize, Eyal Shnarch, Ran Levy, Ben Bogin, Ranit Aharonov, and Noam Slonim. 2020.
\newblock \href {https://doi.org/10.18653/v1/2020.acl-main.523} {Efficient pairwise annotation of argument quality}.
\newblock In \emph{Proceedings of the 58th Annual Meeting of the Association for Computational Linguistics (ACL)}, pages 5772--5781. Association for Computational Linguistics.

\bibitem[{Gorur et~al.(2025)Gorur, Rago, and Toni}]{gorur-etal-2025-large}
Deniz Gorur, Antonio Rago, and Francesca Toni. 2025.
\newblock \href {https://aclanthology.org/2025.coling-main.569/} {Can large language models perform relation-based argument mining?}
\newblock In \emph{Proceedings of the 31st International Conference on Computational Linguistics}, pages 8518--8534, Abu Dhabi, UAE. Association for Computational Linguistics.

\bibitem[{Habernal et~al.(2018)Habernal, Wachsmuth, Gurevych, and Stein}]{habernal2018argument}
Ivan Habernal, Henning Wachsmuth, Iryna Gurevych, and Benno Stein. 2018.
\newblock \href {https://doi.org/10.18653/v1/N18-1185} {The argument reasoning comprehension task: Identification and reconstruction of implicit warrants}.
\newblock In \emph{Proceedings of NAACL-HLT 2018}, pages 1930--1940.

\bibitem[{Hua et~al.(2022)Hua, Yang, and Yang}]{hua-etal-2022-iam}
Xinyu Hua, Zichao Yang, and Yiming Yang. 2022.
\newblock \href {https://doi.org/10.18653/v1/2022.acl-long.146} {Iam: A comprehensive and large-scale dataset for integrated argument mining}.
\newblock In \emph{Proceedings of the 60th Annual Meeting of the Association for Computational Linguistics (Volume 1: Long Papers)}, pages 2052--2064. Association for Computational Linguistics.

\bibitem[{Islam and Goldwasser(2024)}]{islam2024posthoc}
Tunazzina Islam and Dan Goldwasser. 2024.
\newblock \href {https://arxiv.org/abs/2410.05401} {Post-hoc study of climate microtargeting on social media ads with llms: Thematic insights and fairness evaluation}.
\newblock \emph{arXiv preprint arXiv:2410.05401}.
\newblock Preprint.

\bibitem[{Jain et~al.(2024)Jain, Padmanaban, Hazra, Godbole, and Hamann}]{jain2024empowering}
Ayush Jain, Manikandan Padmanaban, Jagabondhu Hazra, Shantanu Godbole, and Hendrik Hamann. 2024.
\newblock \href {https://www.climatechange.ai/papers/iclr2024/27} {Empowering sustainable finance: Leveraging large language models for climate-aware investments}.
\newblock In \emph{ICLR 2024 Workshop on Tackling Climate Change with Machine Learning}. Climate Change AI.

\bibitem[{Krippendorff(2004)}]{krippendorff2004content}
Klaus Krippendorff. 2004.
\newblock \href {https://books.google.com/books/about/Content_Analysis.html?id=q657o3M3C8cC} {\emph{Content Analysis: An Introduction to Its Methodology}}.
\newblock Sage Publications.

\bibitem[{Landis and Koch(1977)}]{landis1977measurement}
J.~Richard Landis and Gary~G. Koch. 1977.
\newblock \href {https://doi.org/10.2307/2529310} {The measurement of observer agreement for categorical data}.
\newblock \emph{Biometrics}, 33(1):159--174.

\bibitem[{Lawrence and Reed(2019)}]{lawrence-reed-2019-survey}
John Lawrence and Chris Reed. 2019.
\newblock \href {https://doi.org/10.1162/coli_a_00364} {Argument mining: A survey}.
\newblock \emph{Computational Linguistics}, 45(4):765--818.

\bibitem[{Lee et~al.(2024)Lee, Peng, Goldberg, Rosenthal, Kotcher, Maibach, and Leiserowitz}]{lee2024can}
Sanguk Lee, Tai-Quan Peng, Matthew~H. Goldberg, Seth~A. Rosenthal, John~E. Kotcher, Edward~W. Maibach, and Anthony Leiserowitz. 2024.
\newblock \href {https://doi.org/10.1371/journal.pclm.0000429} {Can large language models estimate public opinion about global warming? an empirical assessment of algorithmic fidelity and bias}.
\newblock \emph{PLOS Climate}, 3(8):e0000429.

\bibitem[{Leippold(2023)}]{leippold2023thus}
Markus Leippold. 2023.
\newblock Thus spoke gpt-3: Interviewing a large-language model on climate finance.
\newblock \emph{Finance Research Letters}, 53:103617.

\bibitem[{Li et~al.(2025)Li, Schlegel, Sun, Batista-Navarro, and Nenadić}]{li2025largelanguagemodelsargument}
Hao Li, Viktor Schlegel, Yizheng Sun, Riza Batista-Navarro, and Goran Nenadić. 2025.
\newblock \href {http://arxiv.org/abs/2506.16383} {Large language models in argument mining: A survey}.

\bibitem[{Liu et~al.(2021)Liu, Stab, and Gurevych}]{liu-etal-2021-neural-transition}
Yang Liu, Christian Stab, and Iryna Gurevych. 2021.
\newblock \href {https://doi.org/10.18653/v1/2021.acl-long.166} {A neural transition-based model for argumentation mining}.
\newblock In \emph{Proceedings of the 59th Annual Meeting of the Association for Computational Linguistics (ACL)}, pages 2135--2149. Association for Computational Linguistics.

\bibitem[{Liu et~al.(2020)Liu, Zhang, Shang, and Han}]{liu-etal-2020-discourse}
Yang Liu, Sheng Zhang, Jingbo Shang, and Jiawei Han. 2020.
\newblock \href {https://doi.org/10.18653/v1/2020.acl-main.569} {A top-down neural architecture towards text-level parsing of discourse relations}.
\newblock In \emph{Proceedings of the 58th Annual Meeting of the Association for Computational Linguistics (ACL)}, pages 6386--6395. Association for Computational Linguistics.

\bibitem[{MetaAI(2025)}]{meta2025llama4}
MetaAI. 2025.
\newblock {LLaMA} 4: Maverick (17b) — a multimodal mixture-of-experts model.
\newblock Model card / documentation for LLaMA-4 Maverick-17B (128 experts), accessed via ModelScope / model repositories.

\bibitem[{Mistral(2024)}]{mixtral_8x22b}
Mistral. 2024.
\newblock Mixtral 8×22b.
\newblock https://huggingface.co/mistralai/Mixtral-8x22B-v0.1.
\newblock Open-weight mixture-of-experts language model.

\bibitem[{Nisbet(2009)}]{nisbet-2009-frames}
Matthew~C. Nisbet. 2009.
\newblock \href {https://doi.org/10.3200/ENVT.51.2.12-23} {Communicating climate change: Why frames matter for public engagement}.
\newblock \emph{Environment: Science and Policy for Sustainable Development}, 51(2):12--23.

\bibitem[{{OpenAI}(2024)}]{openai2024gpt4o}
{OpenAI}. 2024.
\newblock Gpt-4o system card.
\newblock arXiv preprint arXiv:2410.21276.
\newblock Multimodal “omni” model by OpenAI.

\bibitem[{Otmakhova and Frermann(2025)}]{otmakhova-frermann-2025-narrative}
Yulia Otmakhova and Lea Frermann. 2025.
\newblock \href {https://doi.org/10.18653/v1/2025.findings-acl.477} {Narrative media framing in political discourse}.
\newblock In \emph{Findings of the Association for Computational Linguistics: ACL 2025}, pages 9167--9196, Vienna, Austria. Association for Computational Linguistics.

\bibitem[{Patterson et~al.(2022)Patterson, Gonzalez, H{\"o}lzle, Le, Liang, Munguia, Rothchild, So, Texier, and Dean}]{patterson2022carbon}
David Patterson, Joseph Gonzalez, Urs H{\"o}lzle, Quoc Le, Chen Liang, Lluis-Miquel Munguia, Daniel Rothchild, David So, Maud Texier, and Jeff Dean. 2022.
\newblock \href {https://doi.org/10.1109/MC.2022.3148714} {The carbon footprint of machine learning training will plateau, then shrink}.
\newblock \emph{Computer}, 55(7):18--28.

\bibitem[{Pavlovic and Poesio(2024)}]{pavlovic-poesio-2024-effectiveness}
Maja Pavlovic and Massimo Poesio. 2024.
\newblock \href {https://aclanthology.org/2024.nlperspectives-1.11/} {The effectiveness of llms as annotators: A comparative overview and empirical analysis of direct representation}.
\newblock In \emph{NLPerspectives 2024}. Association for Computational Linguistics.

\bibitem[{Prasad et~al.(2019)Prasad, Webber, Lee, and Joshi}]{prasad-etal-2019-pdtb3}
Rashmi Prasad, Bonnie Webber, Alan Lee, and Aravind Joshi. 2019.
\newblock \href {https://catalog.ldc.upenn.edu/LDC2019T05} {Penn discourse treebank 3.0}.
\newblock Linguistic Data Consortium.

\bibitem[{{ProQuest}(2024)}]{proquest2024factiva}
{ProQuest}. 2024.
\newblock Home - {Dow Jones Factiva} - {LibGuides}.
\newblock \url{https://proquest.libguides.com/factiva}.
\newblock Describes DJID as containing approximately 350,000 classification codes.

\bibitem[{Qwen(2025)}]{qwen3_30b_a3b}
Qwen. 2025.
\newblock Qwen3 30b a3b.
\newblock https://huggingface.co/unsloth/Qwen3-30B-A3B-GGUF.
\newblock Open-weight evaluator model.

\bibitem[{Reimers and Gurevych(2019)}]{reimers-2019-sentence-bert}
Nils Reimers and Iryna Gurevych. 2019.
\newblock \href {https://www.aclweb.org/anthology/D19-1410} {{Sentence-BERT: Sentence Embeddings using Siamese BERT-Networks}}.
\newblock In \emph{Proceedings of the 2019 Conference on Empirical Methods in Natural Language Processing and the 9th International Joint Conference on Natural Language Processing (EMNLP-IJCNLP)}, pages 3982--3992, Hong Kong, China. Association for Computational Linguistics.

\bibitem[{Roberts et~al.(2014)Roberts, Stewart, Tingley, Lucas, Leder-Luis, Gadarian, Albertson, and Rand}]{roberts2014structural}
Margaret~E. Roberts, Brandon~M. Stewart, Dustin Tingley, Christopher Lucas, Jetson Leder-Luis, Sharon Gadarian, Bethany Albertson, and David Rand. 2014.
\newblock \href {https://doi.org/10.1111/ajps.12103} {Structural topic models for open-ended survey responses}.
\newblock \emph{American Journal of Political Science}, 58(4):1064--1082.

\bibitem[{Rousseeuw(1987)}]{rousseeuw-1987-silhouettes}
Peter~J. Rousseeuw. 1987.
\newblock \href {https://doi.org/10.1016/0377-0427(87)90125-7} {Silhouettes: a graphical aid to the interpretation and validation of cluster analysis}.
\newblock \emph{Journal of Computational and Applied Mathematics}, 20:53--65.

\bibitem[{Ruiz-Dolz et~al.(2021)Ruiz-Dolz, Cortes, Garcia, and Garcia-Serrano}]{ruiz-dolz-2021-transformer}
Maria Ruiz-Dolz, Santiago Cortes, Javier Garcia, and Ana Garcia-Serrano. 2021.
\newblock \href {https://doi.org/10.1109/MIS.2021.3072990} {Transformer-based models for automatic identification of argument relations}.
\newblock \emph{IEEE Intelligent Systems}, 36(4):54--62.

\bibitem[{Schirmag et~al.(2025)Schirmag, Wedemeyer, Stechemesser, and Wenz}]{schirmag2025neural}
Tatjana Schirmag, Jakob~H. Wedemeyer, Annika Stechemesser, and Leonie Wenz. 2025.
\newblock \href {https://doi.org/10.1038/s43247-025-02402-1} {Neural topic modeling reveals german television’s climate change coverage}.
\newblock \emph{Communications Earth \& Environment}, 6(1):441.

\bibitem[{Schmid-Petri et~al.(2017{\natexlab{a}})Schmid-Petri, Adam, Schmucki, and Häussler}]{schmid-petri2017}
Hannah Schmid-Petri, Silke Adam, Ivo Schmucki, and Thomas Häussler. 2017{\natexlab{a}}.
\newblock \href {https://doi.org/10.1177/0963662515612277} {A changing climate of skepticism: The factors shaping climate change coverage in the us press}.
\newblock \emph{Public Understanding of Science}, 26(4):498--513.

\bibitem[{Schmid-Petri et~al.(2017{\natexlab{b}})Schmid-Petri, Adam, Schmucki, and Häussler}]{schmidpetri2017}
Hannah Schmid-Petri, Silke Adam, Ivo Schmucki, and Thomas Häussler. 2017{\natexlab{b}}.
\newblock \href {https://doi.org/10.1177/0963662515612277} {A changing climate of skepticism: The factors shaping climate change coverage in the u.s. press}.
\newblock \emph{Public Understanding of Science}, 26(4):498--513.

\bibitem[{Schmid-Petri and Arlt(2016{\natexlab{a}})}]{schmidpetri2016}
Hannah Schmid-Petri and Dorothee Arlt. 2016{\natexlab{a}}.
\newblock \href {https://doi.org/10.1515/commun-2016-0011} {Constructing an illusion of scientific uncertainty? framing climate change in german and british print media}.
\newblock \emph{Communications}, 41(3):265--289.

\bibitem[{Schmid-Petri and Arlt(2016{\natexlab{b}})}]{schmid-petri-2017-climate}
Hannah Schmid-Petri and Dorothee Arlt. 2016{\natexlab{b}}.
\newblock \href {https://doi.org/10.1515/commun-2016-0011} {Constructing an illusion of scientific uncertainty? framing climate change in german and british print media}.
\newblock \emph{Communications}, 41(3):265--289.

\bibitem[{Shiller(2017)}]{shiller2019}
Robert~J. Shiller. 2017.
\newblock \href {https://doi.org/10.1257/aer.107.4.967} {Narrative economics}.
\newblock \emph{American Economic Review}, 107(4):967--1004.

\bibitem[{Stab and Gurevych(2014)}]{stab2014annotating}
Christian Stab and Iryna Gurevych. 2014.
\newblock \href {https://doi.org/10.3115/ablabla} {Annotating argument components and relations in persuasive essays}.
\newblock In \emph{Proceedings of COLING 2014}, pages 1501--1510.

\bibitem[{Stab and Gurevych(2017{\natexlab{a}})}]{stab-gurevych-2017}
Christian Stab and Iryna Gurevych. 2017{\natexlab{a}}.
\newblock \href {https://doi.org/10.18653/v1/N17-1099} {Parsing argumentation structures in persuasive essays}.
\newblock In \emph{Proceedings of the 2017 Conference of the North American Chapter of the Association for Computational Linguistics: Human Language Technologies}, pages 985--995, Los Angeles, California. Association for Computational Linguistics.

\bibitem[{Stab and Gurevych(2017{\natexlab{b}})}]{stefan2017argument}
Christian Stab and Iryna Gurevych. 2017{\natexlab{b}}.
\newblock \href {https://doi.org/10.1162/COLI_a_00295} {Parsing argumentation structures in persuasive essays}.
\newblock In \emph{Computational Linguistics}, volume~43, pages 619--659.

\bibitem[{Tan et~al.(2024)Tan, Li, Wang, Beigi, Jiang, Bhattacharjee, Karami, Li, Cheng, and Liu}]{tan-etal-2024-large}
Zhen Tan, Dawei Li, Song Wang, Alimohammad Beigi, Bohan Jiang, Amrita Bhattacharjee, Mansooreh Karami, Jundong Li, Lu~Cheng, and Huan Liu. 2024.
\newblock \href {https://doi.org/10.18653/v1/2024.emnlp-main.54} {Large language models for data annotation and synthesis: A survey}.
\newblock In \emph{Proceedings of the 2024 Conference on Empirical Methods in Natural Language Processing}, pages 930--957, Miami, Florida, USA. Association for Computational Linguistics.

\bibitem[{Toledo et~al.(2021)Toledo, Kantor, Stanovsky, and Dagan}]{toledo-etal-2021-graph-quality}
Assaf Toledo, Yoav Kantor, Gabriel Stanovsky, and Ido Dagan. 2021.
\newblock \href {https://doi.org/10.26615/978-954-452-072-4_162} {Graph-based argument quality assessment}.
\newblock In \emph{Proceedings of the International Conference on Recent Advances in Natural Language Processing (RANLP 2021)}, pages 1427--1437.

\bibitem[{T{\"o}rnberg(2024)}]{tornberg2024best}
Petter T{\"o}rnberg. 2024.
\newblock \href {https://doi.org/10.48550/arXiv.2402.05129} {Best practices for text annotation with large language models}.
\newblock \emph{arXiv preprint arXiv:2402.05129}.

\bibitem[{Ward(1963)}]{ward-1963-hierarchical}
Joe~H. Ward. 1963.
\newblock \href {https://doi.org/10.1080/01621459.1963.10500845} {Hierarchical grouping to optimize an objective function}.
\newblock \emph{Journal of the American Statistical Association}, 58(301):236--244.

\bibitem[{Webersinke et~al.(2022)Webersinke, Kraus, Bingler, and Leippold}]{webersinke2021climatebert}
Nicolas Webersinke, Mathias Kraus, Julia~Anna Bingler, and Markus Leippold. 2022.
\newblock \href {https://doi.org/10.48550/arXiv.2110.12010} {Climatebert: A pretrained language model for climate-related text}.
\newblock In \emph{Proceedings of the AAAI 2022 Fall Symposium: The Role of AI in Responding to Climate Challenges}.
\newblock ArXiv:2110.12010.

\bibitem[{Wu et~al.(2024)Wu, Masson, Mukherjee, Zhao, Driouich, Xing, Paroubek, and Coustaty}]{masson2024susgengpt}
Qilong Wu, Thibault Masson, Aritra Mukherjee, Yihao Zhao, Chaima Driouich, Xingjian Xing, Patrick Paroubek, and Mickael Coustaty. 2024.
\newblock \href {https://arxiv.org/abs/2412.10906} {Susgen-gpt: A data-centric llm for financial nlp and sustainability report generation}.
\newblock \emph{arXiv preprint arXiv:2412.10906}.
\newblock Preprint.

\bibitem[{Zhuang et~al.(2021)Zhuang, Wayne, Ya, and Jun}]{zhuang-etal-2021-robustly}
Liu Zhuang, Lin Wayne, Shi Ya, and Zhao Jun. 2021.
\newblock \href {https://aclanthology.org/2021.ccl-1.108/} {A robustly optimized {BERT} pre-training approach with post-training}.
\newblock In \emph{Proceedings of the 20th Chinese National Conference on Computational Linguistics}, pages 1218--1227, Huhhot, China. Chinese Information Processing Society of China.

\end{thebibliography}

\appendix
\section{Corpus Curation and Validation}
\subsection{DJID Code Reference}
\label{app:djid}

Table~\ref{tab:djid} lists all Dow Jones Intelligent Identifier (DJID) codes used in our corpus construction, grouped into the three domains introduced in Section~3.1: Core Climate Issues, Energy Transition, and Climate-Affected Sectors.

\begin{table}[h!]
\centering
\small
\begin{tabular}{lll}
\hline
\textbf{Domain} & \textbf{Code} & \textbf{Description} \\
\hline
Core Climate Issues & N/ENV & Environment \\
                    & N/CO2 & Carbon Dioxide / Emissions \\
                    & N/RNW & Renewables \\
\hline
Energy Transition   & N/BFL & Biofuels \\
                    & N/COA & Coal \\
                    & N/NUK & Nuclear Energy \\
                    & N/NGS & Natural Gas \\
\hline
Climate-Affected Sectors & N/AGR & Agriculture \\
                         & N/FST & Forestry \\
\hline
\end{tabular}
\caption{Dow Jones Intelligent Identifier (DJID) codes used in constructing the climate news corpus.}
\label{tab:djid}
\end{table}

\subsection{Validation of DJID Filtering}
\label{app:djid-validation}
To evaluate the reliability of DJID-based filtering, we conducted a two-part validation study.

\paragraph{False-Negative Analysis.}
We sampled 1,000 articles not tagged with climate-related DJIDs and re-screened them using:
\begin{itemize}
    \item \textbf{Keyword Lexicon:} A curated set of 152 climate-related terms (see Appendix~\ref{app:lexicon}) matched against article text.  
    \item \textbf{Supervised Classifier:} A RoBERTa-base model fine-tuned on 5,000 climate vs. non-climate news articles (macro F1 = 0.91). Articles flagged as “climate-related” by either method were manually verified.
\end{itemize}
Out of 1,000 articles, 73 were flagged; manual inspection confirmed 60 as genuinely climate-related. This implies a false-negative rate of $\sim$8\%.

\paragraph{False-Positive Analysis.}
We randomly sampled 500 articles tagged with climate DJIDs. Two annotators independently verified whether each article substantively discussed climate issues (Krippendorff’s $\alpha=0.79$). Only 21 (4.2\%) were deemed false positives, yielding estimated precision $>$95\%.

\paragraph{Summary.}
Overall, DJID filtering achieves recall of $\sim$92\% and precision $>$95\%. While highly accurate, complementary methods (lexicon or classifier-based augmentation) can further enhance coverage in future work.

\subsection{Keyword Lexicon Construction}
\label{app:lexicon}
The climate keyword lexicon was designed to capture articles that may not be tagged with relevant DJIDs. It was constructed from prior climate communication research, IPCC glossaries, and domain-specific terminology in financial reporting. The final lexicon contains 152 terms across five thematic categories. Representative examples are provided in Table~\ref{tab:lexicon}; the full lexicon is released with our supplementary materials for reproducibility.

\begin{table}[t]
\centering
\footnotesize
\begin{tabularx}{\columnwidth}{lX}
\hline
\textbf{Category} & \textbf{Example Keywords} \\
\hline
General Climate & climate change; global warming; greenhouse effect \\
Carbon & carbon; CO2; carbon tax; carbon capture \\
Energy & renewables; solar; wind; fossil fuels; biofuels \\
Finance & ESG; green bonds; carbon markets \\
Policy & Paris Agreement; Kyoto Protocol; net zero \\
\hline
\end{tabularx}
\caption{Representative subsets of the climate keyword lexicon. The full lexicon (152 terms) is provided in the supplementary release.}
\label{tab:lexicon}
\end{table}

\subsection{Preprocessing Details}
\label{app:preproc}

\paragraph{Normalization.}
All articles were standardized through boilerplate removal (e.g., repeated headers, copyright notices), whitespace normalization, and Unicode normalization. Token-level cleaning was avoided to preserve domain-specific terminology.

\paragraph{Near-Duplicate Detection.}
Because wire services frequently release slightly modified repeats of the same story, we applied MinHash-based locality-sensitive hashing (LSH) to identify duplicates. Pairs with Jaccard similarity $\geq 0.9$ were marked as near-duplicates. We retained one canonical version per cluster, removing $\sim$8.7\% of articles. This prevents redundancy and avoids over-weighting syndicated stories.

\paragraph{Final Corpus.}
After preprocessing, the corpus contains 980,061 unique climate-related articles with preserved metadata (timestamps, DJIDs, article source identifiers, etc.), ensuring integrity for diachronic analysis.

\subsection{Corpus Temporal Distribution}
\label{app:volume}

\begin{figure}[h]
\centering
\includegraphics[width=\columnwidth]{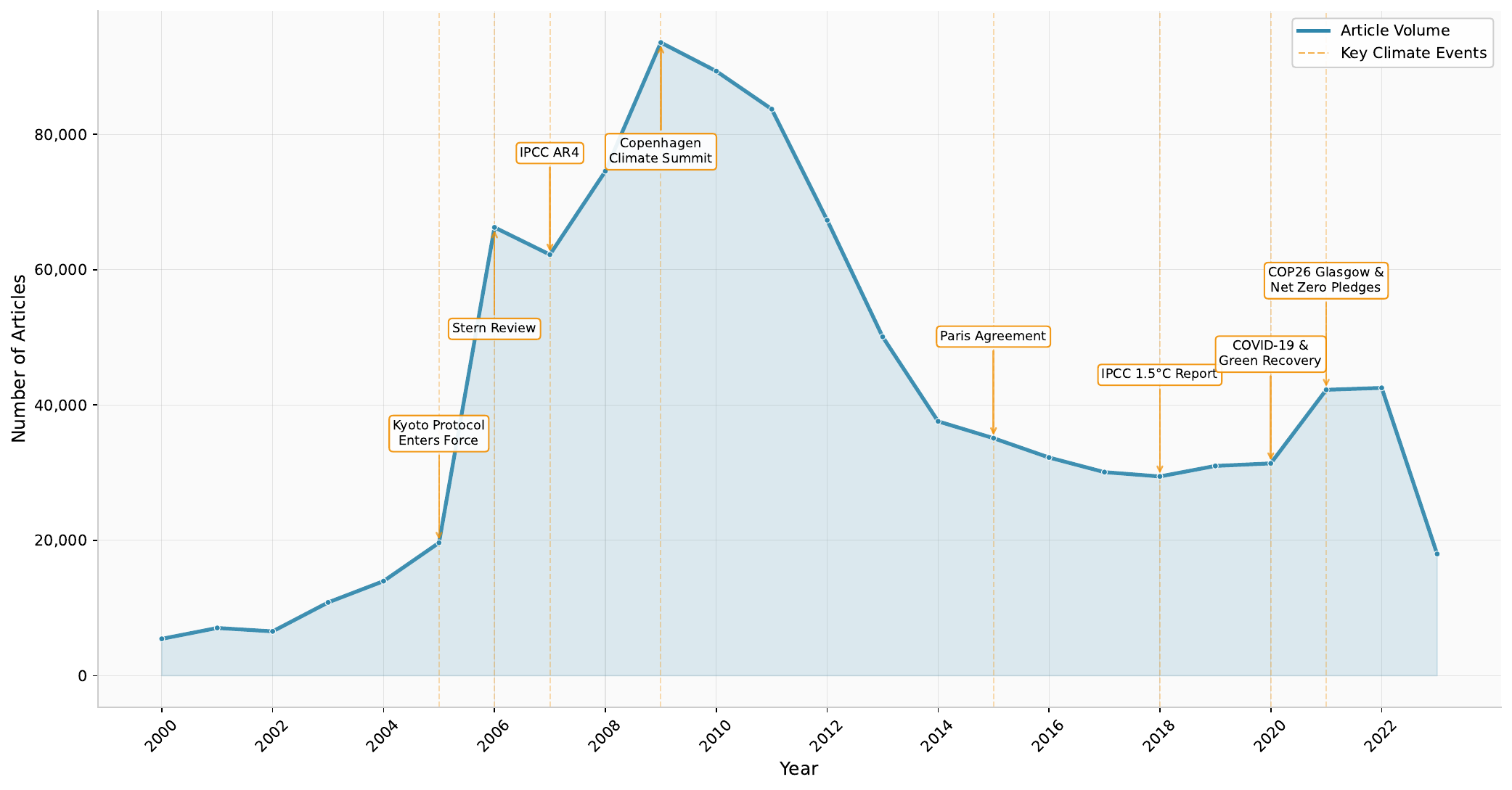} 
\caption{Temporal distribution of Dow Jones climate-related articles (2000–2023), annotated with key climate events.}
\label{volume-appendix}
\end{figure}

\section{Sampling Hyperparameter Analysis}
\label{app:sample}

\subsection{Sensitivity Analysis of MMR \texorpdfstring{$\lambda$}{lambda}}
\begin{figure}[t]
\centering
\begin{tikzpicture}[scale=1, every node/.style={scale=0.90}]
    \begin{axis}[
        width=\columnwidth,
        height=5.2cm,
        xlabel={MMR $\lambda$ (Relevance Weight)},
        ylabel={Macro F1-Score},
        xmin=0.45, xmax=0.95,
        ymin=0.73, ymax=0.76,
        xtick={0.5,0.6,0.7,0.8,0.9},
        ytick={0.73,0.74,0.75,0.76},
        ymajorgrids=true,
        grid style=dashed,
        legend style={
            at={(0.03,0.97)},
            anchor=north west,
            font=\footnotesize,
            draw=none,
            fill=none
        },
        tick label style={font=\scriptsize},
        label style={font=\footnotesize},
        axis y line*=left
    ]
    \addplot[color=blue, mark=*, line width=1pt]
        coordinates {(0.5,0.742) (0.6,0.748) (0.7,0.751) (0.8,0.752) (0.9,0.745)};
    \addlegendentry{Macro F1}
    \end{axis}

    \begin{axis}[
        width=\columnwidth,
        height=5.2cm,
        ylabel={Sample Diversity},
        xmin=0.45, xmax=0.95,
        ymin=0.2, ymax=0.8,
        axis y line*=right,
        axis x line=none,
        ylabel style={yshift=-0.2em, font=\footnotesize},
        tick label style={font=\scriptsize},
    ]
    \addplot[color=red, mark=square*, dashed, line width=1pt]
        coordinates {(0.5,0.68) (0.6,0.61) (0.7,0.52) (0.8,0.41) (0.9,0.29)};
    \addlegendentry{Diversity}
    \end{axis}
\end{tikzpicture}
\vspace{-2mm}
\caption{Trade-off between model performance (Macro F1) and sample diversity across MMR $\lambda$ values.}
\label{fig:lambda_plot}
\vspace{-3mm}
\end{figure}
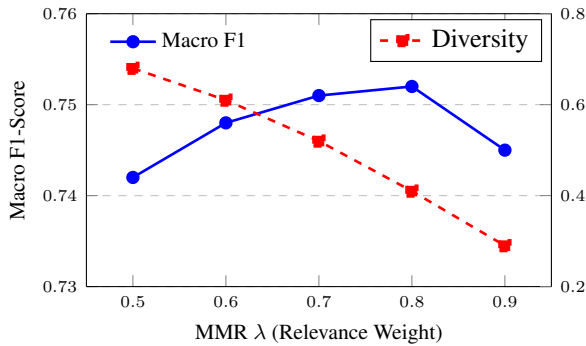

The choice of the Maximal Marginal Relevance (MMR) parameter $\lambda$ governs the trade-off between selecting articles central to a theme (relevance) and those that are novel (diversity). We performed a sensitivity analysis by generating five distinct sub-samples using different $\lambda$ values and evaluating the performance of our downstream argument extraction model on each. As shown in Table~\ref{tab:lambda_sensitivity} and Figure~\ref{fig:lambda_plot}, a value of $\lambda=0.7$ provides a near-optimal balance, achieving high F1-score while retaining substantial sample diversity, measured as the average pairwise cosine distance between selected articles within a cluster. While $\lambda=0.8$ yields a marginal F1-score improvement, this comes at the cost of a significant drop in diversity, making it less suitable for our goal of capturing both mainstream and outlier arguments.

\begin{table}[h!]
\centering
\caption{Performance of the downstream argument extraction model and sample diversity as a function of the MMR parameter $\lambda$. The highest F1-score is in bold.}
\label{tab:lambda_sensitivity}
\resizebox{\columnwidth}{!}{%
\begin{tabular}{@{}lccccc@{}}
\toprule
\textbf{Metric} & $\lambda=0.5$ & $\lambda=0.6$ & $\lambda=0.7$ & $\lambda=0.8$ & $\lambda=0.9$ \\ \midrule
Macro F1 & 0.742 & 0.748 & 0.751 & \textbf{0.752} & 0.745 \\
Precision & 0.739 & 0.745 & 0.750 & 0.758 & 0.761 \\
Recall & 0.745 & 0.751 & 0.752 & 0.746 & 0.730 \\ \midrule
Diversity & 0.68 & 0.61 & 0.52 & 0.41 & 0.29 \\ \bottomrule
\end{tabular}%
}
\end{table}

\subsection{Justification for Thematic Cluster Count (\texorpdfstring{$k$}{k})}
The range for the number of thematic clusters, $k_t$, was constrained to $[5, 15]$ based on qualitative analysis. Table~\ref{tab:k_justification} provides illustrative examples from the 2019-2023 stratum, demonstrating that $k<5$ conflates distinct themes, while $k>15$ creates spurious, overly granular micro-clusters.

\begin{table}[t]
\centering
\caption{Examples of emergent cluster themes at different values of $k$ for the 2019--2023 stratum.}
\label{tab:k_justification}
\resizebox{0.9\columnwidth}{!}{%
\begin{tabular}{@{}ll@{}}
\toprule
\textbf{Value of $k$} & \textbf{Illustrative Cluster Themes} \\ \midrule
$k=4$ & \begin{tabular}[t]{@{}l@{}}- Renewable Energy \& Policy (Conflated) \\ - Corporate ESG \& Finance (Conflated)\end{tabular} \\ \midrule
$k=8$ & \begin{tabular}[t]{@{}l@{}}- Renewable Energy Technology \\ - Climate Finance Policy \\ - Corporate Sustainability Reports \\ - Carbon Markets\end{tabular} \\ \midrule
$k=16$ & \begin{tabular}[t]{@{}l@{}}- Utility-Scale Solar Projects (Redundant) \\ - Onshore Wind Projects (Redundant) \\ - Voluntary Carbon Credits (Redundant)\end{tabular} \\ \bottomrule
\end{tabular}%
}
\vspace{-1mm}
\end{table}

\section{Active Learning Details}
\label{app:AL}

\subsection{Preliminary Model Architecture}
The active learning loop was driven by a preliminary argument component detector based on \textit{RoBERTa-base}. Key training hyperparameters are detailed in Table~\ref{tab:al_model_hparams}. The model achieved a macro F1-score of 79.5\% on a held-out portion of the 1,000-article seed set.

\begin{table}[h!]
\centering
\caption{Hyperparameters for the preliminary argument component detector used in the active learning loop.}
\label{tab:al_model_hparams}
\resizebox{0.8\columnwidth}{!}{%
\begin{tabular}{@{}ll@{}}
\toprule
\textbf{Hyperparameter} & \textbf{Value} \\ \midrule
Base Model & `RoBERTa-base` \\
Learning Rate & 2e-5 \\
Optimizer & AdamW \\
Batch Size & 16 \\
Max Sequence Length & 256 \\
Epochs & 4 \\
Warmup Steps & 100 \\ \bottomrule
\end{tabular}%
}
\end{table}

To clarify the relationship between our various annotated sets:
\begin{enumerate}
    \item \textbf{Seed set for active learning} (1,000 articles): Sampled from full 980k corpus and annotated in-house before final sampling to bootstrap the preliminary argument detector.
    \item \textbf{Final sample} (4,143 articles): Constructed using active learning with the seed-trained detector.
    \item \textbf{Gold standard} (2,000 articles): Subset of the 4,143 sample, annotated via Zooniverse for pipeline evaluation.
    \item \textbf{20k validation corpus} (Appendix D): Separate stratified random sample from the 980k corpus, annotated in-house, used \textit{only} to validate that our 4,143 sample preserves downstream task performance. Not used for training or primary analysis.
\end{enumerate}

\subsection{Active Learning Convergence}
\begin{figure}[h!]
    \centering
    \begin{tikzpicture}
        \begin{axis}[
            width=\columnwidth,
            height=6cm,
            xlabel={Active Learning Iteration},
            ylabel={JSD of Entropy Distributions},
            xmin=0, xmax=5,
            ymin=0, ymax=0.3,
            xtick={1,2,3,4},
            legend pos=north east,
            ymajorgrids=true,
            grid style=dashed,
            legend style={font=\footnotesize},
        ]
        \addplot[color=blue, mark=*, line width=1.5pt] coordinates { (1,0.24) (2,0.11) (3,0.06) (4,0.04) };
        \addplot[color=red, dashed, domain=0:5, samples=2] {0.05};
        \legend{JSD, Convergence Threshold}
        \end{axis}
    \end{tikzpicture}
    \caption{Convergence of the active learning loop. The JSD between selected article sets drops below the 0.05 threshold at iteration 4.}
    \label{fig:al_convergence}
\end{figure}
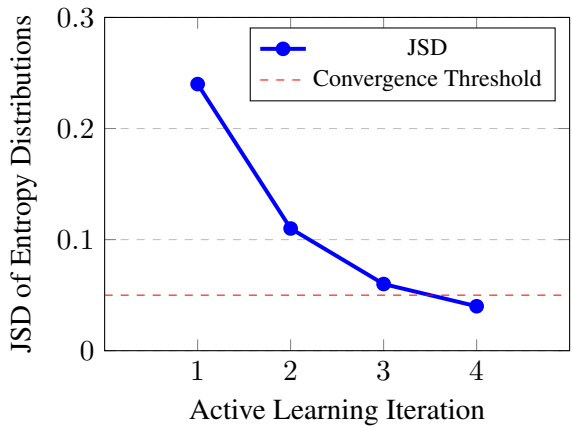

The active learning process converged after 4 iterations. We defined convergence as the point where the Jensen-Shannon divergence (JSD) between the entropy distributions of articles selected in consecutive iterations fell below a threshold of 0.05. Figure~\ref{fig:al_convergence} plots this trend, demonstrating that the set of "hardest" articles identified by the model stabilized quickly.

\section{Extended Validation Results}
\label{app:validation}

To verify the integrity of our 4,143-article sample, we conducted a rigorous validation protocol focused on two key criteria: (1) the preservation of downstream task performance and (2) the fidelity of metadata distributions compared to the parent corpus.

\subsection{Downstream Task Performance Preservation}
The most critical test of a sub-corpus is its utility for the intended downstream task. We evaluated this by training an argument extraction model on our sample and comparing its performance against the same model architecture trained on a much larger 20,000-article stratified random sample.

\paragraph{Model Architecture.} The argument extraction model is a token classifier built upon \texttt{RoBERTa-large}. We added a linear layer on top of the final hidden states of the RoBERTa model to classify each token into IOB format (Inside, Outside, Beginning) for our target components (`Claim`, `Premise`). The model was fine-tuned end-to-end. Key hyperparameters, chosen via a limited search on a development set, are detailed in Table~\ref{tab:validation_model_hparams}.

\begin{table}[h!]
\centering
\caption{Hyperparameters for the \texttt{RoBERTa-large} argument extraction model used in our validation experiment.}
\label{tab:validation_model_hparams}
\resizebox{0.9\columnwidth}{!}{%
\begin{tabular}{@{}ll@{}}
\toprule
\textbf{Hyperparameter} & \textbf{Value} \\ \midrule
Base Model & \texttt{roberta-large} \\
Learning Rate & 1e-5 \\
Optimizer & AdamW with linear decay \\
Weight Decay & 0.01 \\
Batch Size & 8 \\
Max Sequence Length & 512 tokens \\
Training Epochs & 5 \\ \bottomrule
\end{tabular}%
}
\end{table}

\paragraph{Results.} As shown in Table~\ref{tab:validation_perf}, the model trained on our 4,143-article sample achieved a macro F1-score of 75.1\%. This represents a performance degradation of only 3.3 percentage points compared to the 78.4\% F1-score from the model trained on the 20k-article baseline. This result is highly encouraging, demonstrating that our sampling strategy preserves over 95\% of the performance while using less than 21\% of the training data, thus confirming its high data efficiency.

\begin{table}[h!]
\centering
\caption{Detailed performance comparison for the argument extraction task. Both models were evaluated on the same held-out test set of 1,500 articles. The best scores for each metric are in bold.}
\label{tab:validation_perf}
\resizebox{\columnwidth}{!}{%

\begin{tabular}{@{}lccccccc@{}}
\toprule
\multicolumn{1}{c}{\textbf{Training Sample}} & \multicolumn{3}{c}{\textbf{Claim}} & \multicolumn{3}{c}{\textbf{Premise}} & \textbf{Macro} \\
\cmidrule(lr){2-4} \cmidrule(lr){5-7} \cmidrule(lr){8-8}
 & P & R & F1 & P & R & F1 & F1 \\ \midrule
20k Stratified Random & \textbf{0.801} & \textbf{0.753} & \textbf{0.776} & \textbf{0.769} & \textbf{0.824} & \textbf{0.795} & \textbf{0.784} \\
\textbf{Our Sample (4k)} & 0.782 & 0.725 & 0.752 & 0.723 & 0.781 & 0.751 & 0.751 \\ \bottomrule
\end{tabular}%
}
\end{table}

\subsection{Distributional Similarity Fidelity}
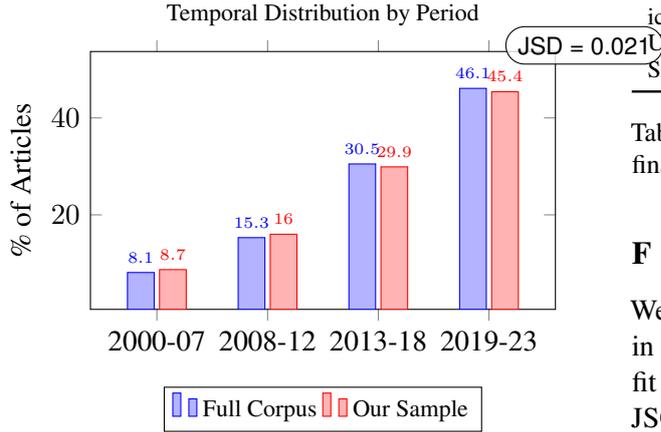
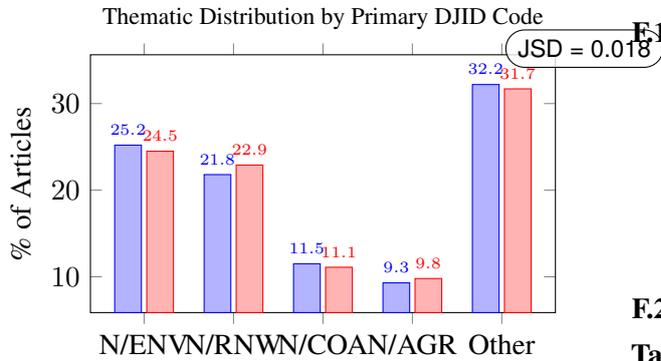
\begin{figure}[h!]
    \centering
    \begin{subfigure}{\columnwidth}
        \begin{tikzpicture}
            \begin{axis}[
                width=\linewidth,
                height=5cm,
                ybar,
                enlargelimits=0.2,
                title style={font=\small},
                title={Temporal Distribution by Period},
                ylabel={\% of Articles},
                symbolic x coords={2000-07, 2008-12, 2013-18, 2019-23},
                xtick=data,
                nodes near coords,
                nodes near coords style={font=\tiny, /pgf/number format/fixed, /pgf/number format/precision=1},
                nodes near coords align={vertical},
                bar width=10pt,
                legend style={at={(0.5,-0.3)}, anchor=north, legend columns=-1, font=\footnotesize},
            ]
            \addplot coordinates {(2000-07, 8.1) (2008-12, 15.3) (2013-18, 30.5) (2019-23, 46.1)};
            \addplot coordinates {(2000-07, 8.7) (2008-12, 16.0) (2013-18, 29.9) (2019-23, 45.4)};
            \legend{Full Corpus, Our Sample}
            \end{axis}
            \node[draw, rounded rectangle, fill=white, fill opacity=0.8, text opacity=1, font=\sffamily\footnotesize] at (6.5,3.5) {JSD = 0.021};
        \end{tikzpicture}
        \caption{Comparison of temporal distributions.}
        \label{fig:dist_year}
    \end{subfigure}
    
    \vspace{0.5cm} 
    
    \begin{subfigure}{\columnwidth}
        \begin{tikzpicture}
            \begin{axis}[
                width=\linewidth,
                height=5cm,
                ybar,
                enlargelimits=0.15,
                title style={font=\small},
                title={Thematic Distribution by Primary DJID Code},
                ylabel={\% of Articles},
                symbolic x coords={N/ENV, N/RNW, N/COA, N/AGR, Other},
                xtick=data,
                nodes near coords,
                nodes near coords style={font=\tiny, /pgf/number format/fixed, /pgf/number format/precision=1},
                nodes near coords align={vertical},
                bar width=10pt,
                legend style={at={(0.5,-0.3)}, anchor=north, legend columns=-1, font=\footnotesize},
            ]
            \addplot coordinates {(N/ENV, 25.2) (N/RNW, 21.8) (N/COA, 11.5) (N/AGR, 9.3) (Other, 32.2)};
            \addplot coordinates {(N/ENV, 24.5) (N/RNW, 22.9) (N/COA, 11.1) (N/AGR, 9.8) (Other, 31.7)};
            \end{axis}
            \node[draw, rounded rectangle, fill=white, fill opacity=0.8, text opacity=1, font=\sffamily\footnotesize] at (6.5,3.5) {JSD = 0.018};
        \end{tikzpicture}
        \caption{Comparison of thematic distributions.}
        \label{fig:dist_topic}
    \end{subfigure}
    \caption{Comparison of key metadata distributions between the full 980k-article corpus and our 4k-article sample, confirming high distributional fidelity.}
    \label{fig:dist_sim}
\end{figure}

We further validated that our sample's metadata distribution faithfully mirrors that of the parent 980k-article corpus. We measured the Jensen-Shannon divergence (JSD) for key distributions, where a lower JSD score indicates higher similarity. Figure~\ref{fig:dist_sim} provides a visual comparison for temporal and thematic distributions, confirming a very high degree of fidelity (JSD < 0.03 for both).

\section{Frame Typology}
\label{app:frames}

Our analysis employs a predefined eight-frame typology for climate discourse in financial news. This typology synthesizes categories from communication studies \citep{nisbet-2009-frames, schmid-petri-2017-climate} with finance-specific frames (e.g., economic risk, market dynamics) observed in preliminary corpus exploration. Table~\ref{tab:frames-appendix} provides definitions of each frame.

\begin{table}[h!]
\centering
\small
\begin{tabular}{p{0.28\linewidth} p{0.65\linewidth}}
\toprule
\textbf{Frame} & \textbf{Definition} \\
\midrule
Economic Opportunity & Frames climate change as growth, innovation, and investment potential. \\
Economic Risk & Highlights financial losses, stranded assets, and risks to firms or markets. \\
Regulatory Compliance & Focuses on laws, policies, and regulatory burdens or incentives. \\
Technological Solution & Emphasizes innovation, R\&D, and technical fixes to climate challenges. \\
Environmental Urgency & Stresses ecological severity and the need for rapid action. \\
Social Responsibility & Invokes ethics, corporate responsibility, and societal expectations. \\
Market Dynamics & Frames climate in terms of competition, supply-demand, and positioning. \\
Uncertainty Skepticism & Expresses doubt or skepticism about climate science, policies, or impacts. \\
\bottomrule
\end{tabular}
\caption{Eight-frame typology for climate discourse in financial news.}
\label{tab:frames-appendix}
\end{table}

\section{Prompt Templates and Schemas}
\label{app:prompts}

We release all prompt templates and schemas used in the Actor--Frame--Argument (AFA) pipeline. To fit the two-column format, we present compact JSON schemas and keep explanatory text brief. Full JSON schema files and scripts are included in our supplementary materials.

\subsection{General Protocol}
\begin{itemize}\setlength\itemsep{0pt}
  \item Output must be a valid JSON object (no prose).
  \item Missing fields: return \texttt{null} or empty list.
  \item Frames must match exactly with the predefined typologies.
\end{itemize}

\subsection{Stage 1: Actor--Stance}
\textbf{Task:} Extract actors, type, stance, supporting quote.  
\textbf{Schema:}
\begin{verbatim}
{
 "actors": [{
   "name": "string",
   "actor_type": "company|gov|indiv|org",
   "stance": "supportive|opposing|neutral|mixed",
   "quote_text": "string",
   "climate_relevance": "string"
 }]
}
\end{verbatim}

\subsection{Stage 2: Frame Classification}
\textbf{Task:} Assign one primary and optional secondary frame.  
\textbf{Schema:}
\begin{verbatim}
{
 "primary_frame": "frame_enum",
 "secondary_frame": "frame_enum_or_null",
 "justification": "string", 
 "climate_connection": "string"
}
\end{verbatim}

\subsection{Stage 3: Argument Extraction}
\textbf{Task:} Extract claim, evidence, warrant, impact, and optional supporting arguments.  
\textbf{Schema:}
\begin{verbatim}
{
 "claim": "string",
 "evidence": ["string", "..."],
 "warrant": "string",
 "impact": "string",
 "supporting_arguments": [{
   "claim": "string",
   "evidence": ["string", "..."],
   "warrant": "string"
 }]
}
\end{verbatim}

\subsection{DVF Judge Prompts}
\textbf{Dimensions:} Completeness, Faithfulness, Coherence, Climate Relevance.  
Each atomic check is scored on a continuous scale in $[0,1]$, where $0$ indicates complete failure, $1$ indicates full success, and intermediate values capture partial correctness.  

\textbf{Schema (example):}
\begin{verbatim}
{
 "completeness": {"actors":0.0-1.0,
                  "stance":0.0-1.0,
                  "frames":0.0-1.0,
                  "arguments":0.0-1.0},
 "faithfulness": {"quote_alignment":0.0-1.0,
                  "para_equivalence":0.0-1.0},
 "coherence": {"links_preserved":0.0-1.0,
               "schema_wellformed":0.0-1.0},
 "climate_relevance": {"on_topic":0.0-1.0,
                       "peripheral_excluded":0.0-1.0}
}
\end{verbatim}

\subsection{Inference Settings}
Temperature = 0.2, Top-$p$ = 0.9, max tokens = 512 (stages 1--2), 768 (stage 3), 256 (DVF). Stop sequences: \texttt{\detokenize{\{"\{\n","```\}}}. Outputs validated against schemas; retries on failure.

\section{Human-Annotated Gold Standard}
\label{app:gold}

To establish a reliable evaluation anchor for our AFA pipeline, we constructed a gold-standard dataset of 2,000 articles sampled from the corpus. This section provides full details of the annotation protocol, guidelines, adjudication process, and agreement metrics.

\subsection{Annotator Training and Background}
Annotations were collected via \textit{Zooniverse}, a widely used citizen science platform for distributed human annotation. Instead of a small team of in-house coders, we leveraged a large pool of volunteers. To ensure annotation quality, we implemented the following measures:

\begin{itemize}\setlength\itemsep{0pt}
  \item \textbf{Onboarding Tutorial:} All contributors completed an interactive tutorial with examples of actor, frame, and argument annotations, as well as practice tasks with feedback.
  \item \textbf{Redundancy:} Each article was annotated independently by at least $k=5$ volunteers to mitigate individual errors.
  \item \textbf{Aggregation:} Final labels were assigned via majority vote (for categorical tasks such as stance and frames) or consensus heuristics (for argument spans, using token-level agreement).

\end{itemize}

While annotators did not have formal training in linguistics or climate communication, the redundancy and aggregation protocol yielded high inter-annotator agreement (see Table~\ref{tab:iaa}). This approach demonstrates the feasibility of scalable citizen science annotation for complex discourse tasks.

\subsection{Annotation Guidelines}
The schema covered all components of the Actor--Frame--Argument (AFA) pipeline:
\begin{itemize}\setlength\itemsep{0pt}
  \item \textbf{Actor--Stance}: identify all actors (company, government, NGO, individual) and classify stance as \textit{supportive}, \textit{opposing}, \textit{neutral}, or \textit{mixed}. Each actor annotation required a supporting quote.
  \item \textbf{Frame Classification}: assign one \textit{primary frame} and (if present) one \textit{secondary frame} from the eight-frame typology (see Appendix~\ref{app:frames}). Annotators provided a brief justification.
  \item \textbf{Argument Extraction}: decompose argument structure into (i) central claim, (ii) evidence spans, (iii) warrant linking claim and evidence, and (iv) impact (stated or implied). Additional supporting arguments were annotated if present.
\end{itemize}
Examples of annotated documents and excerpts from the guidelines are released with our supplementary materials.

\subsection{Annotation Procedure}
Each of the 2,000 articles was annotated on Zooniverse by at least five independent contributors. For categorical tasks (stance, frame), final labels were determined by majority vote. For span-based tasks (actors, evidence, warrants), we aggregated annotations using token-level agreement and retained spans marked by at least 60\% of contributors. Disagreements were resolved through consensus aggregation rather than individual adjudication. This redundancy protocol provides a robust approximation of expert annotation quality while leveraging the scale of citizen science.

\section{Annotation Guidelines (Excerpt)}
\label{app:guidelines}

This section provides an excerpt from the full annotation guidelines used for training human coders. These examples illustrate how annotators distinguished labels and handled common edge cases in the Actor--Frame--Argument (AFA) schema. The complete guidelines and codebook are released with our supplementary materials.

\subsection{Actor--Stance Annotation}
\textbf{Task:} Identify actors (organization, company, government, NGO, or individual) and assign stance.  

\textbf{Rules:}
\begin{itemize}\setlength\itemsep{0pt}
  \item An actor must be explicitly mentioned in the text. Generic terms such as “analysts” or “critics” are excluded unless tied to a named entity.
  \item If multiple subsidiaries or affiliates are mentioned, treat them as separate actors only if they present distinct stances.
  \item Stance definitions:
    \begin{itemize}
      \item \textbf{Supportive:} Explicitly endorses climate action, investment, or regulatory compliance.
      \item \textbf{Opposing:} Rejects or resists climate measures, or highlights negative impacts.
      \item \textbf{Neutral:} Mentions climate issues without evaluative judgment.
      \item \textbf{Mixed:} Expresses both supportive and opposing positions within the same context.
    \end{itemize}
  \item Each stance annotation must be supported by a verbatim quote.
\end{itemize}

\textbf{Example:}  
*“ExxonMobil said new carbon rules will increase costs for energy producers.”*  
→ Actor: ExxonMobil; Stance: Opposing; Quote: verbatim.

\subsection{Frame Classification}
\textbf{Task:} Assign one primary frame and, if present, one secondary frame from the eight-frame typology.  

\textbf{Rules:}
\begin{itemize}\setlength\itemsep{0pt}
  \item Primary frame = the dominant way the issue is presented.  
  \item Secondary frame = optional, used only if the text clearly invokes a secondary dimension.  
  \item If multiple frames appear, select the one most central to the claim or argument as primary.  
\end{itemize}

\textbf{Example:}  
*“Investors see renewable energy as the next big growth opportunity.”*  
→ Primary frame: Economic Opportunity; Secondary frame: Technological Solution.  

\subsection{Argument Extraction}
\textbf{Task:} Decompose into claim, evidence, warrant, and impact.  

\textbf{Rules:}
\begin{itemize}\setlength\itemsep{0pt}
  \item \textbf{Claim:} Main proposition advanced by the actor.  
  \item \textbf{Evidence:} Factual or statistical support cited in the text. Use verbatim snippets.  
  \item \textbf{Warrant:} The reasoning linking claim and evidence. This can be implicit but should be paraphrased.  
  \item \textbf{Impact:} Stated or implied consequence if the claim is accepted.  
\end{itemize}

\textbf{Example:}  
Sentence: “BlackRock argues that green portfolios outperform traditional funds, with ESG funds gaining 4.3\% in 2020.”
\begin{itemize}
  \item Claim: “Green portfolios outperform traditional funds.”  
  \item Evidence: “ESG funds gaining 4.3\% in 2020.”  
  \item Warrant: “Financial returns justify sustainable investment.”   
\end{itemize}

\subsection{Ambiguity Resolution}
\begin{itemize}\setlength\itemsep{0pt}
  \item If a passage fits multiple plausible labels, select the one best supported by evidence.  
  \item Use “Mixed” stance only if a single actor explicitly expresses both support and opposition.  
  \item Annotators flagged unclear cases for adjudication rather than guessing.  
\end{itemize}

\subsection{Inter-Annotator Agreement}
Agreement was computed separately for each component type, using measures appropriate to multi-annotator settings:

\begin{itemize}\setlength\itemsep{0pt}
  \item \textbf{Actor Identification}: token-level precision, recall, and F1 computed pairwise across annotators, averaged with macro-F1.
  \item \textbf{Stance Classification}: Fleiss’ $\kappa$ and Krippendorff’s $\alpha$ (nominal scale) over four stance categories.
  \item \textbf{Frame Assignment}: Krippendorff’s $\alpha$ on the eight-frame set for primary frames; Jaccard similarity averaged across annotator pairs for secondary frames.
  \item \textbf{Argument Components}: span-level F1 (claims, evidence, warrants, impacts), aggregated by majority vote. Macro-F1 reported across components.
\end{itemize}

\subsection{Consensus Derivation for Span-Based Tasks}
\label{app:span_consensus}

For span-based tasks (actors, arguments), we derive a single gold-standard annotation from multiple human labels using a two-step token-aggregation and span-reconstruction
procedure. This approach consolidates fine-grained token agreement while avoiding over-fragmentation of long argument components.

\paragraph{Step 1: Token-Level Agreement.}
Let each article be tokenized as a sequence
$T = [t_1, t_2, \dots, t_m]$.
Each annotator $a_i$ provides a binary label sequence
$y_i(t_j) \in \{0, 1\}$, where $1$ indicates that token $t_j$
belongs to a span of interest (e.g., actor mention, claim, evidence).
For each token, we compute an agreement score:
\[
A(t_j) = \frac{1}{N} \sum_{i=1}^{N} y_i(t_j),
\]
where $N$ is the number of annotators (typically $N=5$).
Tokens satisfying $A(t_j) \ge \tau$ (with $\tau = 0.5$)
are retained, producing a consensus mask
$M = [m_1, m_2, \dots, m_m]$ where $m_j = 1$
iff $A(t_j) \ge \tau$.

\paragraph{Step 2: Span Reconstruction.}
Contiguous retained tokens are merged into unified spans:
\[
S_k = \{t_p, \dots, t_q\} \text{ such that } m_p = \dots = m_q = 1, 
\; (m_{p-1}, m_{q+1}) = 0.
\]
If multiple annotator spans overlap semantically,
we compute pairwise span-level $\mathrm{F1}$:
\[
\mathrm{F1}(H_i, S_k) =
\frac{2 \times |\mathrm{Tokens}(H_i) \cap \mathrm{Tokens}(S_k)|}
{|\mathrm{Tokens}(H_i)| + |\mathrm{Tokens}(S_k)|},
\]
where $H_i$ is a human span. Spans with $\mathrm{F1} > 0.7$
are merged by maximal coverage (minimal start, maximal end).

\paragraph{Handling Non-Span Tokens.}
Tokens not marked by any annotator (e.g., function words)
are assigned the outside label $O$:
\[
y_i(t_j) = 0 \quad \forall i, \text{ if } t_j \text{ unmarked.}
\]
They are excluded from aggregation but retained for evaluation.
Token-level $\mathrm{F1}$ is computed as
\[
\mathrm{F1} = \frac{2 \times \mathrm{TP}}
{2 \times \mathrm{TP} + \mathrm{FP} + \mathrm{FN}},
\]
where consistently labeled $O$ tokens contribute neither to the
numerator nor denominator.

\paragraph{Illustrative Example.}
Sentence:
\emph{``BlackRock stated that climate-focused portfolios will drive growth.''}

\noindent Three annotators mark:
\begin{itemize}
    \item A$_1$: [BlackRock]; [climate-focused portfolios will drive growth]
    \item A$_2$: [BlackRock Inc.]; [climate-focused portfolios will drive growth]
    \item A$_3$: [BlackRock]; [portfolios will drive growth]
\end{itemize}
After token alignment and majority voting, only
\emph{BlackRock} and \emph{climate-focused portfolios will drive growth}
reach consensus. All other tokens remain labeled $O$.
This yields clean, non-overlapping gold spans.

\paragraph{Algorithmic Summary (Pseudo-code).}
\begin{verbatim}
Input: tokenized text T, annotation matrix Y, threshold tau
for each token t_j in T:
    A(t_j) = mean(Y[:, j])
    M[j] = 1 if A(t_j) >= tau else 0
merge contiguous tokens where M[j] = 1 into spans S_k
for each human span H_i:
    compute F1(H_i, S_k)
merge overlapping spans if F1 > 0.7 using maximal coverage
return consensus spans S
\end{verbatim}

This two-step method preserves boundary fidelity for short mentions
while maintaining robustness for longer, clause-level arguments,
yielding a reproducible gold standard for span-based evaluation.

\subsection{Use in Evaluation}
The adjudicated consensus labels form the gold standard for all intrinsic evaluations in Section~5. Model outputs are evaluated against this set using precision, recall, and F1. In addition, the gold standard provides the validation baseline for the Decompositional Verification Framework (DVF), ensuring that automatic verification scores are anchored to reliable human annotations.

\section{Decompositional Verification Framework}
\subsection{Decompositional Verification Sub-Checks}
\label{app:dvfsub}

DVF decomposes evaluation into atomic sub-checks that are easier for models to verify and humans to audit.

\begin{itemize}
  \item \textbf{Completeness:} 
    (1) Are all actors identified? 
    (2) Is stance extracted? 
    (3) Are frames assigned? 
    (4) Are argument structures fully captured?
  \item \textbf{Faithfulness:} 
    (1) Does each extracted component align with a direct quote? 
    (2) Is paraphrase semantically equivalent?
  \item \textbf{Structural Coherence:} 
    (1) Are actor–frame–argument links preserved? 
    (2) Is the schema well-formed?
  \item \textbf{Climate Relevance:} 
    (1) Is the extracted frame genuinely about climate? 
    (2) Are peripheral issues (e.g., generic market news) excluded?
\end{itemize}

\subsection{Decompositional Verification Framework Results}
\label{app:dvf}

This appendix provides the full breakdown of DVF evaluations, complementing the aggregate results reported in Section~\ref{subsec:validation}. Recall that DVF evaluates extractions across four dimensions: \textit{completeness}, \textit{faithfulness}, \textit{structural coherence}, and \textit{climate relevance}. Scores are averaged over a 2,000-article gold-standard set, with validation against an independent 500-sample human evaluation.

\subsection{Per-Judge Performance}
DVF employs four distinct judge models to mitigate self-grading bias: GPT-4o (primary), Claude-Sonnet-4, Qwen3-30B A3B, and Mixtral-8$\times$22B. Table~\ref{tab:dvf_judges} reports per-judge scores before aggregation. GPT-4o and Claude achieve the highest consistency, while open-weight models provide competitive but slightly noisier estimates, ensuring robustness via cross-family triangulation.

\begin{table}[t]
\centering
\small
\setlength{\tabcolsep}{3.5pt} 
\begin{tabular}{lcccc}
\toprule
\textbf{Dim.} & \textbf{GPT-4o} & \textbf{Claude} & \textbf{Qwen3} & \textbf{Mixtral} \\
\midrule
Completeness & 0.842 & 0.837 & 0.814 & 0.808 \\
Faithfulness & 0.902 & 0.895 & 0.871 & 0.868 \\
Coherence    & 0.801 & 0.794 & 0.772 & 0.769 \\
Relevance    & 0.869 & 0.861 & 0.842 & 0.836 \\
\bottomrule
\end{tabular}
\caption{DVF dimension-level scores by individual judge, averaged over the 2k-article gold standard.}
\label{tab:dvf_judges}
\end{table}

\subsection{Human Validation Study}
To anchor automated DVF scores, we constructed a separate human evaluation set of 500 randomly sampled model outputs. Each output was annotated by three trained coders using the DVF rubric, independent from the gold-standard annotators. Agreement across coders was high: $\kappa = 0.78$ (faithfulness), $\alpha = 0.74$ (completeness), and $\alpha = 0.71$ (coherence), indicating substantial reliability.  

Table~\ref{tab:dvf_human} compares aggregated model-judge scores against human annotations. Automated DVF evaluations track human ratings closely, with deviations under 0.03 across dimensions. This validation confirms that DVF provides a faithful proxy for expert assessment while scaling efficiently.

\begin{table}[h]
\centering
\small
\begin{tabular}{lcc}
\toprule
\textbf{Dimension} & \textbf{Human Score} & \textbf{DVF (avg.)} \\
\midrule
Completeness & 0.842 & 0.821 \\
Faithfulness & 0.895 & 0.857 \\
Coherence   & 0.781 & 0.792 \\
Relevance   & 0.864 & 0.816 \\
\bottomrule
\end{tabular}
\caption{Comparison of human-coded DVF scores and aggregated automated DVF scores over the 500-item validation set.}
\label{tab:dvf_human}
\end{table}

\noindent These results demonstrate that while judge-specific variation exists, the aggregate DVF scores remain well-validated against human judgment, justifying their use as the main evaluation metric in the body of the paper.

\section{Results}
\subsection{Actor Category Shares}
\label{app:actor_table}

Table~\ref{tab:actor_shares} reports the exact proportions of each actor 
category across the four temporal strata, complementing the trend 
visualization in Figure~\ref{fig:actor_trends}.

\begin{table}[h]
\centering
\small
\begin{tabular}{lccccc}
\toprule
\textbf{Period} & \textbf{Comp.} & \textbf{Fin. Inst.} & \textbf{Gov.} & \textbf{NGO} & \textbf{Indiv.} \\
\midrule
2000--2007 & 20.7 & 14.8 & 31.4 & 19.2 & 13.9 \\
2008--2012 & 24.7 & 16.3 & 28.1 & 12.4 & 18.5 \\
2013--2018 & 26.5 & 24.2 & 22.7 & 11.1 & 15.5 \\
2019--2023 & 27.8 & 33.6 & 19.6 & 8.7  & 10.3 \\
\bottomrule
\end{tabular}
\caption{Actor category shares (\%) across temporal strata.}
\label{tab:actor_shares}
\end{table}

\subsection{Frame Distributions Over Time}
\label{app:frame_table}

Table~\ref{tab:frame_shifts} reports exact frame proportions (\%) in early (2000--2014) vs.\ recent (2019--2023) periods, with significance tests for temporal independence. These values complement Figure~\ref{fig:frame_trends} and the changepoint analysis described in Section~\ref{subsec:transformation}.

\begin{table}[h]
\centering
\small
\begin{tabular}{lccc}
\toprule
\textbf{Frame} & \textbf{2000--2007} & \textbf{2019--2023} & \textbf{$\Delta$} \\
\midrule
Economic Opportunity   & 18.3 & 33.7 & +15.4 \\
Economic Risk          & 31.2 & 18.1 & $-13.1$ \\
Regulatory Compliance  & 26.8 & 16.9 & $-9.9$ \\
Technological Solution & 15.2 & 21.4 & +6.2 \\
Market Dynamics        & 11.8 & 14.3 & +2.5 \\
Environmental Urgency  & 9.2  & 8.9  & $-0.3$ \\
Social Responsibility  & 6.2  & 7.8  & +1.6 \\
Uncertainty/Skepticism & 8.7  & 4.9  & $-3.8$ \\
\bottomrule
\end{tabular}
\caption{Frame distribution shift between early (2000--2007) and recent (2019--2023) periods (percentages). Chi-square tests indicate significant temporal change for most frames (*** $p{<}.001$, ** $p{<}.01$, * $p{<}.05$; n.s.\ not significant), matching the main-text trend narrative.}
\label{tab:frame_shifts}
\end{table}

\subsection{Actor–Frame Association Matrix}
\label{app:actor_frame}

Table~\ref{tab:actor_frame_matrix} reports standardized residuals from a $\chi^{2}$ 
test of independence between actor groups and frame usage. Positive values indicate 
over-representation; negative values indicate under-representation. Significance levels 
are adjusted with Bonferroni correction.

\begin{table}[t]
\centering
\scriptsize
\setlength{\tabcolsep}{3pt} 
\begin{tabular}{lcccc}
\toprule
\textbf{Actor Type} & \textbf{Econ.\ Opp.} & \textbf{Econ.\ Risk} & \textbf{Tech.\ Sol.} & \textbf{Env.\ Urg.} \\
\midrule
Companies & $+4.2^{***}$ & $-2.8^{**}$ & $+2.1^{*}$ & $-3.9^{***}$ \\
Financial Inst. & $+5.7^{***}$ & $-1.9^{*}$ & $+1.6$ & $-4.2^{***}$ \\
Govt./Regulators & $+0.8$ & $+1.2$ & $-0.6$ & $+1.8^{*}$ \\
NGOs/Advocacy & $-5.1^{***}$ & $+0.9$ & $-1.3$ & $+6.8^{***}$ \\
Researchers/Experts & $-1.4$ & $+2.3^{*}$ & $+3.4^{***}$ & $+2.7^{**}$ \\
\bottomrule
\end{tabular}
\vspace{-1mm}
\caption{Actor–frame association matrix (standardized residuals from $\chi^{2}$ test). 
Positive values indicate over-representation. Significance: *** $p{<}.001$, ** $p{<}.01$, * $p{<}.05$.}
\label{tab:actor_frame_matrix}
\vspace{-2mm}
\end{table}

\subsection{Qualitative Error Analysis}
\label{app:error-analysis}

To better understand the model’s limitations, we manually examined 150 randomly sampled mispredictions across all Actor–Frame–Argument (AFA) components. 
Each case was annotated by two authors following the same schema used in the main evaluation. 
Three dominant error categories emerged, together explaining over 80\% of observed failures.

\paragraph{Actor Ambiguity (32\%).}
Ambiguities often arise in passages quoting multiple entities or spokespersons. 
For instance:

\begin{quote}
\small
``BlackRock and several market analysts said green bonds will outperform traditional fixed income.''
\end{quote}

The model frequently merges ``BlackRock'' and ``analysts'' into a single actor span or assigns a collective ``financial institutions'' type. 
Disentangling speaker roles may require discourse-level coreference or dependency cues.

\paragraph{Frame Overlap (27\%).}
When both opportunity and technological themes are expressed, the system struggles to assign a dominant frame. 
Example:

\begin{quote}
\small
``Investments in renewable innovation secure competitiveness for firms adapting to net-zero policies.''
\end{quote}

Here, human annotators favored \emph{Economic Opportunity} as primary and \emph{Technological Solution} as secondary, whereas the model reversed them. 
Introducing hierarchical or multi-label frame modeling could mitigate this confusion.

\paragraph{Argument Boundary Drift (25\%).}
In complex sentences with subordinate clauses, claim boundaries tend to over-extend. 
Example:

\begin{quote}
\small
``Experts argue that stricter disclosure rules, which may initially burden firms, will ultimately enhance transparency and investor confidence.''
\end{quote}

The model extracted the entire sentence as a claim, omitting separation between claim and evidence. 
Span-level attention or discourse segmentation could help constrain extraction.

\paragraph{Less Frequent Errors.}
Other minor issues include stance misclassification for sarcastic tone (8\%) and actor-type misidentification for supranational organizations such as the IMF or World Bank (6\%). 

\paragraph{Implications.}
Overall, qualitative inspection indicates that most errors stem from rhetorical or syntactic complexity rather than lexical gaps. 
These findings suggest that future work should combine LLM annotation with discourse-aware parsers and fine-grained validation to better handle ambiguity in financial commentary.

\end{document}